\pdfoutput=1
\documentclass{JHEP3}
\usepackage{amsmath}
\usepackage{graphics}
\usepackage{epsfig}
\usepackage{url}

\def\({\left(}
\def\){\right)}
\def\[{\left[}
\def\]{\right]}
\def\<{\langle}
\def\>{\rangle}
\def\ap{\alpha}
\def\bt{\beta}
\def\gm{\gamma}
\def\de{\delta}
\def\ep{\varepsilon}
\def\zt{\zeta}
\def\sg{\sigma}

\newcommand{\be}{\begin{equation}}
\newcommand{\ee}{\end{equation}}
\newcommand{\bal}{\begin{aligned}}
\newcommand{\eal}{\end{aligned}}

\newcommand{\labell}[1]{\label{#1}}

\title{All-loop Mondrian Diagrammatics and 4-particle Amplituhedron}

\author{Yang An$^a$, Yi Li$^a$, Zhinan Li$^a$,
Junjie Rao$^{ab}$\footnote{Corresponding author: jrao@aei.mpg.de}\\
{$^a$Zhejiang Institute of Modern Physics, Zhejiang University, Hangzhou, 310027, P. R. China}\\
{$^b$Max Planck Institute for Gravitational Physics (Albert Einstein Institute), 14476 Potsdam, Germany}}

\abstract{Based on 1712.09990 which handles the 4-particle amplituhedron at 3-loop, we have found an extremely
simple pattern, yet far more non-trivial than one might naturally expect: the all-loop Mondrian diagrammatics.
By further simplifying and rephrasing the key relation of positivity in the amplituhedron setting,
remarkably, we find a completeness relation unifying all diagrams of the Mondrian types for the 4-particle integrand
of planar $\mathcal{N}\!=\!4$ SYM to all loop orders, each of which can be mapped to a simple
product following a few plain rules designed for this relation. The explicit examples we investigate span from
3-loop to 7-loop order, and based on them, we classify the basic patterns of Mondrian diagrams into four types:
the ladder, cross, brick-wall and spiral patterns. Interestingly, for some special combinations of ordered subspaces
(a concept defined in the previous work), we find failed exceptions of the completeness relation
which are called ``anomalies'', nevertheless,
they substantially give hints on the all-loop recursive proof of this relation. These investigations are closely
related to the combinatoric knowledge of separable permutations and Schr\"{o}der numbers, and go even further
from a diagrammatic perspective. For physical relevance, we need to further consider dual conformal invariance for two
basic diagrammatic patterns to correct the numerator for a local integrand involving one or both of such patterns, while the
denominator encoding its pole structure and also the sign factor, are already fixed by rules of the completeness relation.
With this extra treatment to ensure the integrals are dual conformally invariant, each Mondrian diagram can be exactly
translated to its corresponding physical loop integrand after being summed over all ordered subspaces that admit it.}

\keywords{Amplitudes, Loop integrands}

\begin{document}
\maketitle

\section{Introduction}

The amplituhedron proposal for 4-particle all-loop integrand of planar $\mathcal{N}\!=\!4$ SYM
is a novel reformulation which only
uses positivity conditions for all physical poles to construct the loop integrand.
From \cite{Arkani-Hamed:2013jha,Arkani-Hamed:2013kca} we have understood the 2-loop case,
then in a previous work \cite{Rao:2017fqc}, we have further proved the 3-loop case.
For the latter, even though a brute-force calculation suffices,
we prefer to take the advantage of the Mondrian diagrammatic interpretation to trivialize it significantly.

An unexplained fact of this interpretation is, we do not know why the ``spurious terms'' which have no
Mondrian diagrammatic meaning will finally sum to zero. As far as the 3-loop case this simply works, but
beyond 3-loop there is no guarantee as the straightforward check will be hardly feasible within the current framework.
Furthermore, non-Mondrian diagrams also contribute to the planar 4-particle integrand,
so there might be one possibility that, beyond 3-loop, the failure of cancelation between spurious terms can justify the
existence of non-Mondrian diagrams. However, we will not explore this direction for now, but content ourselves with
a less involved goal: to understand the Mondrian diagrammatics to all loop orders and see to what extend it may aid
the techniques of amplituhedron.

Immediately at 2-loop order, the only positive quantity is
\be
D_{12}=X_{12}+Y_{12},~X_{12}\equiv(x_2-x_1)(z_1-z_2),~Y_{12}\equiv(y_2-y_1)(w_1-w_2), \labell{eq-2}
\ee
here we have chosen to work in the \textit{ordered subspace} $X(12)Z(21)Y(12)W(21)$ or more briefly $X(12)Y(12)$, since
we will always assume the orderings of $z$ and $w$ are opposite to those of $x$ and $y$ respectively.
In this setting, $X_{12}$ and $Y_{12}$ are always positive since $x_1\!<\!x_2$, $z_2\!<\!z_1$ and $y_1\!<\!y_2$,
$w_2\!<\!w_1$.

\begin{figure}
\begin{center}
\includegraphics[width=0.35\textwidth]{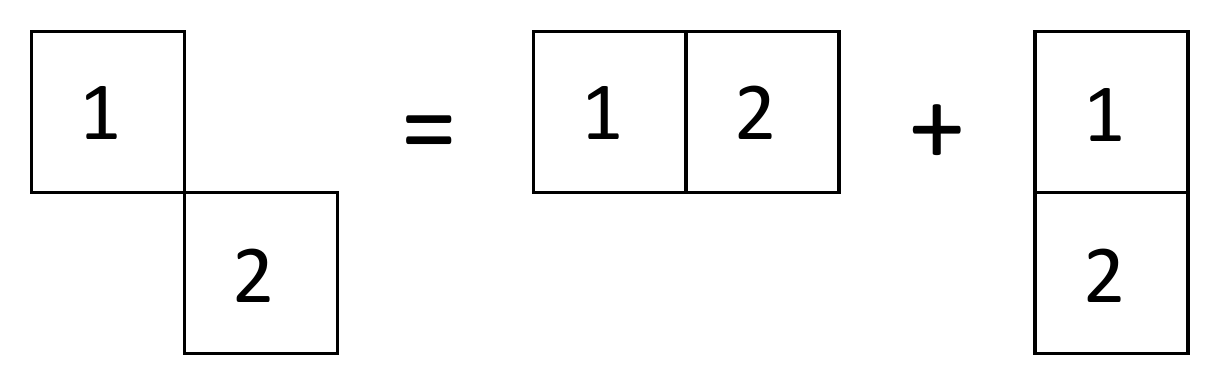}
\caption{Diagrammatic identity $D_{12}\!=\!X_{12}\!+\!Y_{12}$ in subspace $X(12)Y(12)$.} \label{fig-4}
\end{center}
\end{figure}

\begin{figure}
\begin{center}
\includegraphics[width=0.9\textwidth]{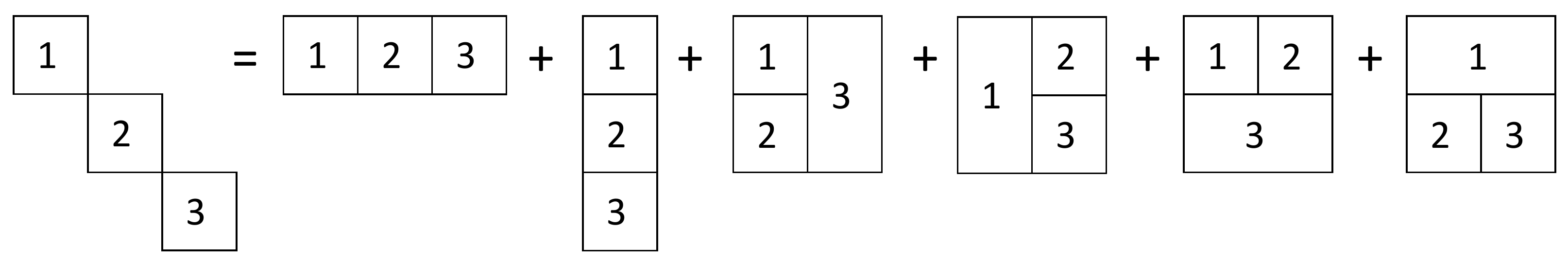}
\caption{Completeness relation at 3-loop in subspace $X(123)Y(123)$.} \label{fig-5}
\end{center}
\end{figure}

Recall that for a Mondrian diagram, all internal lines can be oriented either horizontally or vertically
as borders of adjacent loops, and consequently only 3- and 4-vertex are admitted.
Inside it, any two loops may only have a horizontal contact, vertical contact or no contact,
which leads to plain rules for mapping the diagram to its Mondrian factor, given by (assuming $x_i\!<\!x_j$, $y_i\!<\!y_j$)
\be
\bal
\textrm{horizontal contact: }&X_{ij}\\
\textrm{vertical contact: }&Y_{ij}\\
\textrm{no contact: }&D_{ij}\!=\!X_{ij}\!+\!Y_{ij}~(\textrm{always taking }i\!<\!j\textrm{ for }D_{ij}) \labell{eq-1}
\eal
\ee
between two loops labelled by $i,j$. For the 2-loop example above, relation $D_{12}\!=\!X_{12}\!+\!Y_{12}$
can be directly interpreted as a diagrammatic identity given in figure \ref{fig-4}. Of course, the LHS diagram
is not legal by itself but it is easy to imagine that at 3-loop or higher, two loops inside a Mondrian diagram
do not necessarily have a contact. Explicitly, at 3-loop we have a more interesting relation
\be
D_{12}D_{13}D_{23}=X_{12}X_{23}D_{13}+Y_{12}Y_{23}D_{13}
+X_{13}X_{23}Y_{12}+X_{12}X_{13}Y_{23}+X_{12}Y_{13}Y_{23}+Y_{12}Y_{13}X_{23},
\ee
for the ordered subspace $X(123)Y(123)$, again it can be diagrammatically interpreted as an identity
given in figure \ref{fig-5}. The six terms in its RHS are Mondrian diagrams of all possible orientations
in this subspace, which include two ladders and four tennis courts.

For the 2-loop and 3-loop cases, if we go back to the original definitions, namely
$X_{ij}\!=\!(x_j\!-\!x_i)(z_i\!-\!z_j)$ and $Y_{ij}\!=\!(y_j\!-\!y_i)(w_i\!-\!w_j)$, we can simply sum each
so-called \textit{seed diagram} given above over all subspaces of $x,z,y,w$ that admit it,
and then obtain the correct numerator of its corresponding actual integrand,
as elaborated in the previous work. So far, the 4-particle amplituhedron up to 3-loop is fully understood.

In general, Mondrian diagrams of higher loop orders satisfy this neat pattern: the product of all $D_{ij}$'s
can be expanded as a sum of all topologies of all possible orientations for a particular ordered subspace.
A primitive way to prove this \textit{completeness relation} is to enumerate all relevant diagrams then simplify their sum.
For each distinct topology of a specific orientation, there is only one consistent way to arrange the numbers
inside its boxes as one can easily verify with diagrams above. For example, to fill three numbers into the third diagram
in the RHS of figure \ref{fig-5}, of which the expression by contact rules is $X_{13}X_{23}Y_{12}$, it is trivial
to first put number 3 in its rightmost box since $x_1\!<\!x_2\!<\!x_3$ (recall the increasing direction of $x$ is leftward),
then put numbers 1 and 2 vertically in its left obeying $y_1\!<\!y_2\!<\!y_3$ (the increasing direction of $y$
is downward). In fact, this diagram only implies $x_1,x_2\!<\!x_3$ and $y_1\!<\!y_2$, and $X(123)Y(123)$ is one of
the ordered subspace it admits. The extra information of orderings $x_1\!<\!x_2$ and $y_1,y_2\!<\!y_3$ is
irrelevant for this diagram, since boxes $1,2$ have a vertical contact so that we can only compare $y_1,y_2$ as
they are ``parallel'' in the $x$ direction. And boxes $1,2$ together can be treated as a larger box which has a
horizontal contact with box 3, then the same logic applies. As mentioned in the previous work, this diagram corresponds to
subspace $X(\sg(12)\,3)Y(12)$, where $\sg(12)$ includes two permutations of loop numbers $1,2$.

One may doubt whether there is anything special of subspace $X(12)Y(12)$ at 2-loop or $X(123)Y(123)$ at 3-loop,
such that the completeness relation must hold. To check this, we have worked out all the other orderings,
namely $Y(21)$ fixing $X(12)$ at 2-loop and $Y(132),Y(213),Y(231),Y(312),Y(321)$ fixing $X(123)$ at 3-loop, and
the completeness relation holds with no exception. Since it is the relative ordering between $x$- and $y$-space
that really matters, it is convenient to fix $X(12\ldots L)$ and focus on $Y(\sg_1\sg_2\ldots\sg_L)$.

However, immediately at 4-loop, we find an exception $Y(2413)$ (and $Y(3142)$ as its reverse) which will be named as
the ``anomaly''. The combinatorics accounting for such a failure is,
$(2413)$ is a \textit{non-separable} permutation of $(1234)$.
Literally, it means there is no way to chop it into two sub-permutations such that any number in one subset is always
larger (or smaller) than any number in the other. For $(2413)$, we have $(2)(413)$, $(24)(13)$ and $(241)(3)$, but none of
these is separable, since the in-between numbers $2,3$ are now enclosing the boundary numbers $1,4$.
Beyond 4-loop, the anomalies will occur even more frequently, and we understand why there is no problem with the 2-loop
or 3-loop case: its cardinality is too small to form a non-separable permutation.
The concept of separable permutations will turn out to be closely connected to
the key relation $D\!=\!X\!+\!Y$ inspired by amplituhedron, in the proof of the completeness relation.

\begin{figure}
\begin{center}
\includegraphics[width=0.4\textwidth]{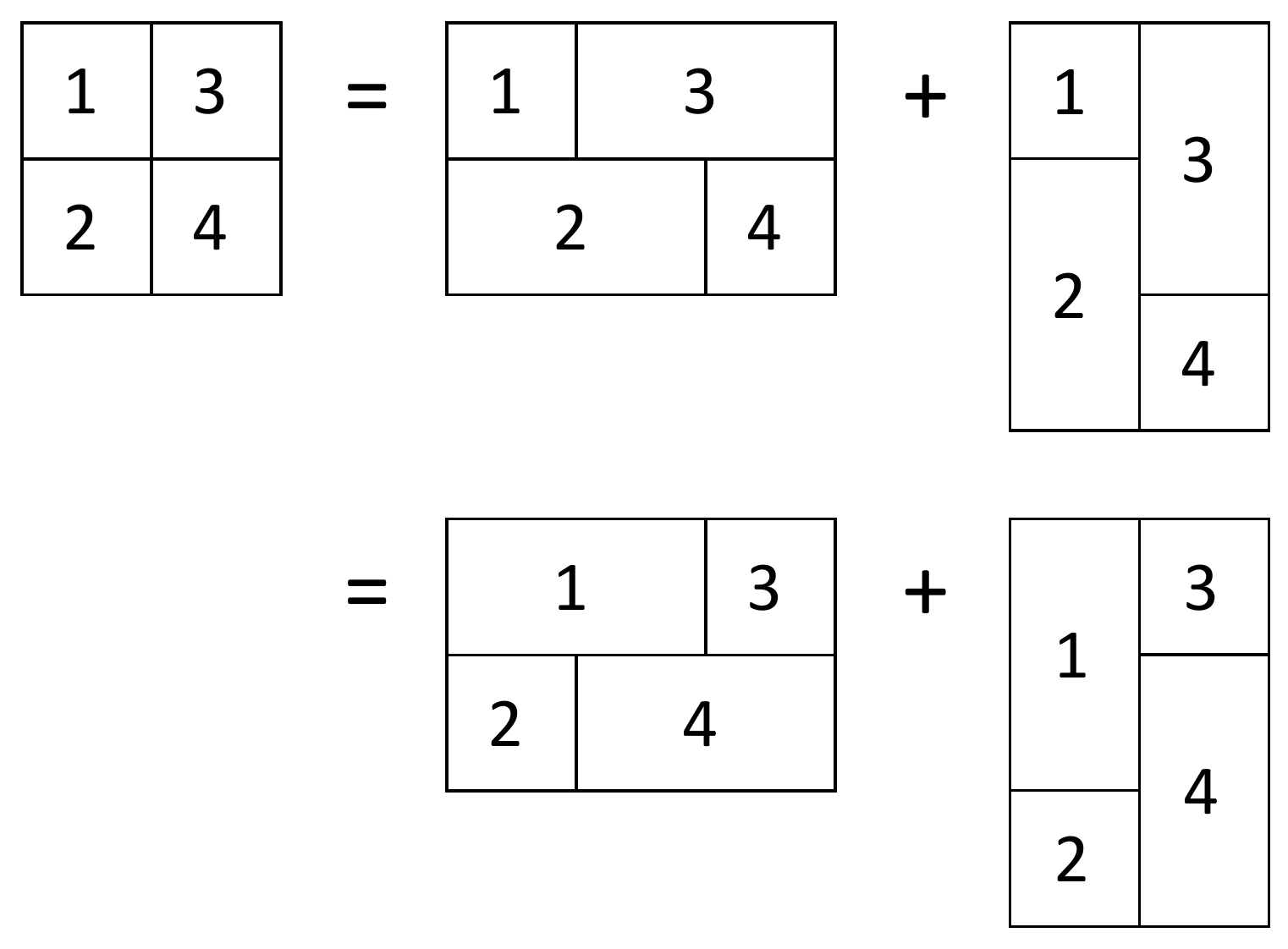}
\caption{Identity connecting a cross diagram and a pair of brick-wall diagrams
at 4-loop in subspace $Y(1324)$.} \label{fig-6}
\end{center}
\end{figure}

Another new feature at 4-loop is, we now have a ``cross'' topology of which for the first time the sign factor is $-1$,
instead of $+1$ for all others! This minus sign in fact ensures the cancelation between a cross diagram and other two
derivative ``brick-wall'' diagrams. Conversely, we can use $D_{ij}\!=\!X_{ij}\!+\!Y_{ij}$ to stretch a 4-vertex
into a vertical and a horizontal contact as demonstrated in figure \ref{fig-6}, where the relevant relation
is $D_{23}\!=\!Y_{32}\!+\!X_{23}$ or $D_{14}\!=\!Y_{14}\!+\!X_{14}$, in subspace $Y(1324)$ for example.

\begin{figure}
\begin{center}
\includegraphics[width=0.35\textwidth]{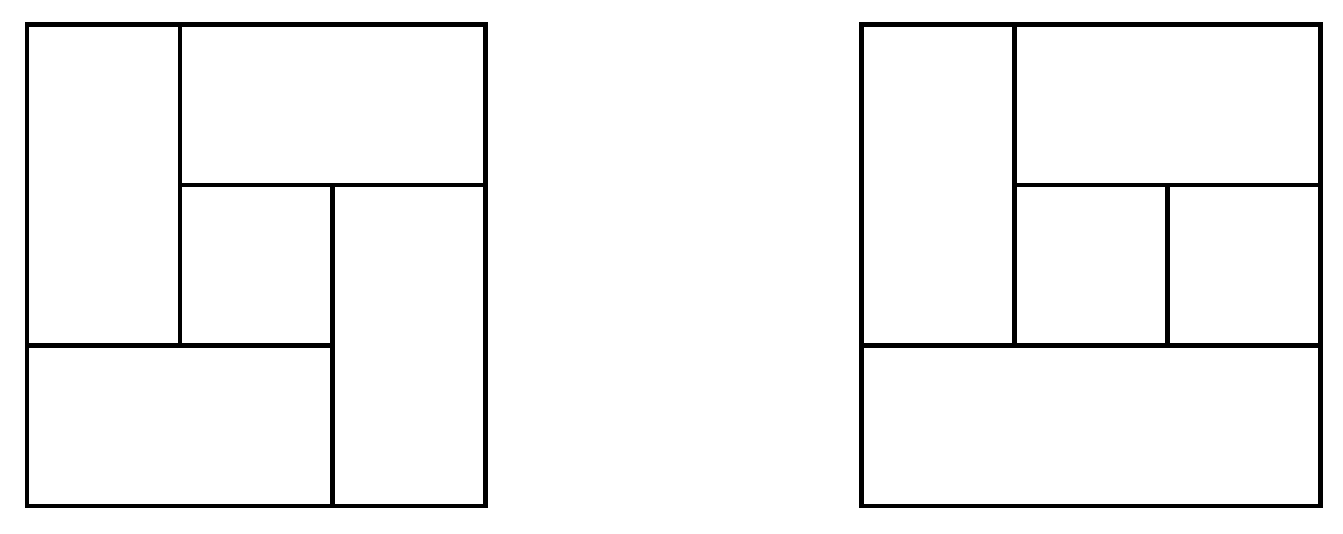}
\caption{A pinwheel (spiral) diagram and a generalized ladder diagram at 5-loop.} \label{fig-7}
\end{center}
\end{figure}

\begin{figure}
\begin{center}
\includegraphics[width=0.6\textwidth]{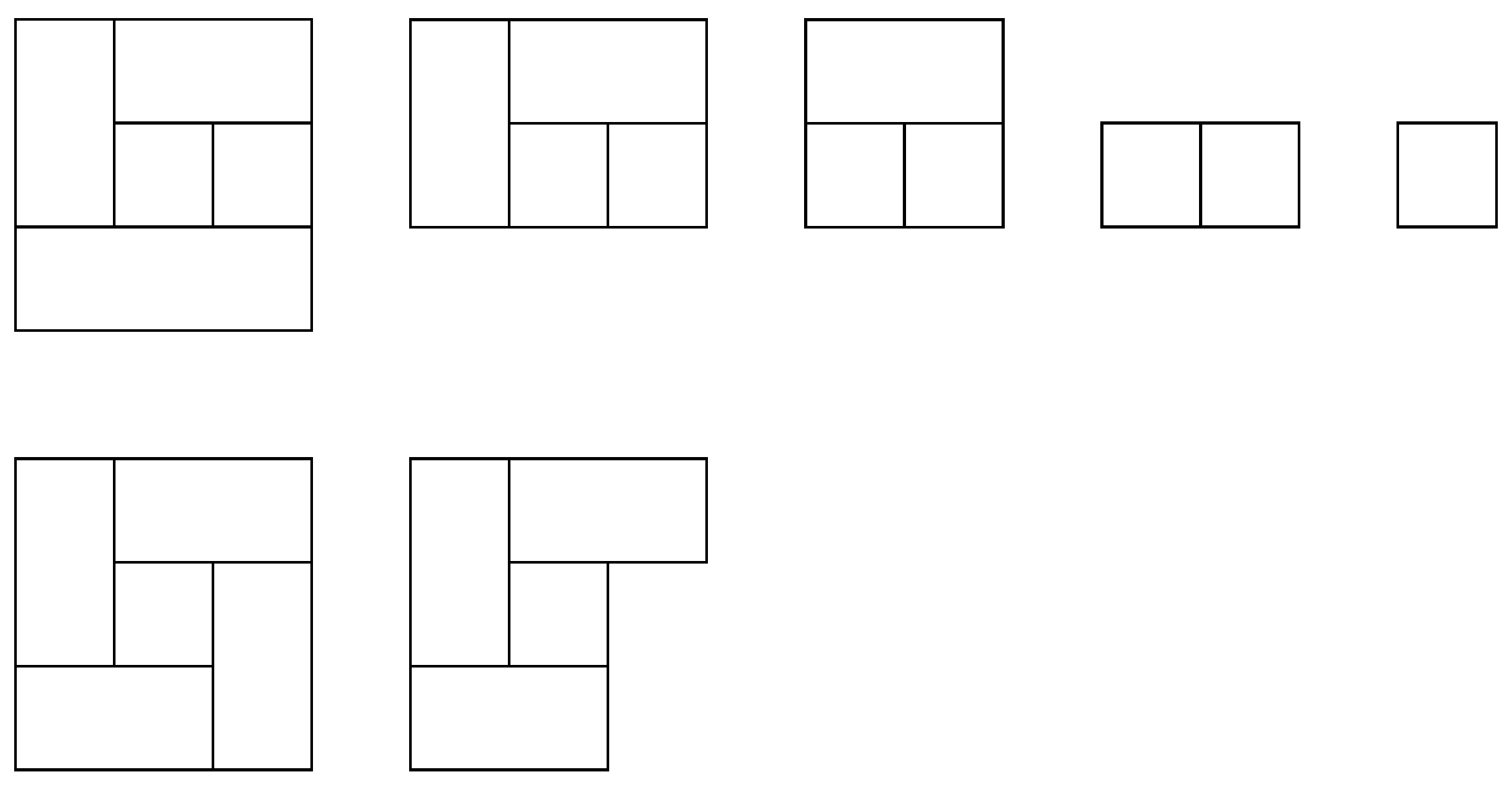}
\caption{A crucial difference between a generalized ladder and a pinwheel.} \label{fig-8}
\end{center}
\end{figure}

Besides the cross and brick-wall topologies, at 5-loop we also encounter the ``spiral'' pattern
of which the simplest form is a ``pinwheel'' diagram given in the left of figure \ref{fig-7}.
In general, if a Mondrian diagram contains none of the cross, brick-wall and spiral patterns,
it will be categorized as the generalized ladder, such as the one in the right of figure \ref{fig-7}.
Although they look very alike, there is a crucial difference: we can detach boxes from a generalized ladder
one by one, while maintaining its exterior profile as a box at each step. In contrast, this is not possible
for the pinwheel, as demonstrated in figure \ref{fig-8}. This difference in fact will lead to a significant combinatoric
distinction for arranging the numbers in their boxes.

In summary, starting from 5-loop, we have seen all of the four basic patterns of Mondrian topologies: the generalized
\textbf{L}adder as the simplest type, the \textbf{C}ross and its derivative \textbf{B}rick-wall, and the \textbf{S}piral.
Note that these patterns are compatible which means one Mondrian diagram can contain all of them. Since the L-pattern
is more trivial than the others, we will not explicitly indicate it as the non-trivial types are only restricted to
the C-, B- and S-pattern. As we will see, the most non-trivial one, namely the S-pattern, can
significantly complicate separable permutations and the corresponding anomalies.

The anomalies which begin to show up at 4-loop, appear to be a nuisance of the perfect completeness relation.
But the fact is the other way around: it is the analysis of anomalies that sheds light on the proof of
completeness relation by recursion. Only for separable permutations the relation $D_{ij}\!=\!X_{ij}\!+\!Y_{ij}$
can lead to the completeness relation, after it is delicately interpreted as a recursive implication.

Understanding all these diagrammatic machineries is almost sufficient for us to translate the relevant Mondrian diagrams
into dual conformally invariant integrals of physical interest. Yet we need to manually heal dual conformal invariance
for two basic diagrammatic patterns so that they get the correct integrand numerators, which are a special L-pattern
and the S-pattern. It is known that merely a dual conformally invariant integrand is not sufficient,
but the integral has to be convergent under certain off-shell limits as well, which imposes more strict constraints
on the numerators.

This article is organized as follows. Section \ref{sec2} investigates the Mondrian diagrammatics, and its connection
to dual conformally invariant local integrands at 4-loop in different ordered subspaces, then studies
the anomaly which first shows up at 4-loop. Section \ref{sec3} does the same for the 5-loop case, in which we also encounter
the new spiral pattern, and there are more concrete examples of the cancelation between crosses and brick-walls,
as well as more anomalies which help us further understand non-separable permutations.
Section \ref{sec4} presents the all-loop recursive proof of the completeness relation for an arbitrary separable permutation
by delicately separating all relevant Mondrian diagrams into the contributing and the cancelling parts.
The former is sufficient for the completeness relation and we can choose the subspace $Y(12\ldots L)$ to maximally
simplify the proof without loss of generality, while any extra complexity will be taken care by the analysis of
cancelation for the latter.

\newpage
\section{Mondrian Diagrammatics at 4-loop}
\label{sec2}

Now, we start the investigation of Mondrian diagrammatics at 4-loop, in two ordered subspaces $Y(1234)$ and $Y(1324)$,
by using the completeness relation that can help us to enumerate all Mondrian topologies of all possible orientations.
After this we also elaborate how to obtain a physical integrand from a Mondrian diagram by summing the latter over
all ordered subspaces that admit it, while the additional consideration of dual conformal invariance
is needed for one special pattern.
The completeness relation fails in subspace $Y(2413)$ since $(2413)$ is a non-separable permutation of $(1234)$,
and this exception leads to an anomaly.

\subsection{Ordered subspace $X(1234)Y(1234)$}

The 4-particle integrand of planar $\mathcal{N}\!=\!4$ SYM is known up to 10-loop
\cite{Bern:2005iz,Bern:2006ew,Bern:2007ct,Eden:2012tu,Bourjaily:2011hi,Bourjaily:2015bpz,Bourjaily:2016evz}
but the 4-particle amplituhedron agrees with these results only up to 3-loop so far. Although we would like to push ahead,
a brute-force calculation for the 4-loop amplituhedron is not realistic yet. For the ordered subspaces of
$y$ and $w$ there are already $4!\!\times\!4!\!=\!576$ combinations, let alone to sum it over
$4!\!=\!24$ permutations of four loop numbers. Even though with the insight of Mondrian diagrammatics,
we still have to take care of the spurious terms, as what the previous work does, plus there is a new non-Mondrian
topology at 4-loop to be accounted for. At last, the result of amplituhedron still must be prudently
separated as a sum of local integrands. While in contrast, the Mondrian diagrams are
local by themselves after we sum them over all ordered subspaces.
Therefore, it is better to directly start with the enumeration of Mondrian diagrams, at the cost of
setting aside the non-Mondrian contributions for the moment.

\begin{figure}
\begin{center}
\includegraphics[width=0.6\textwidth]{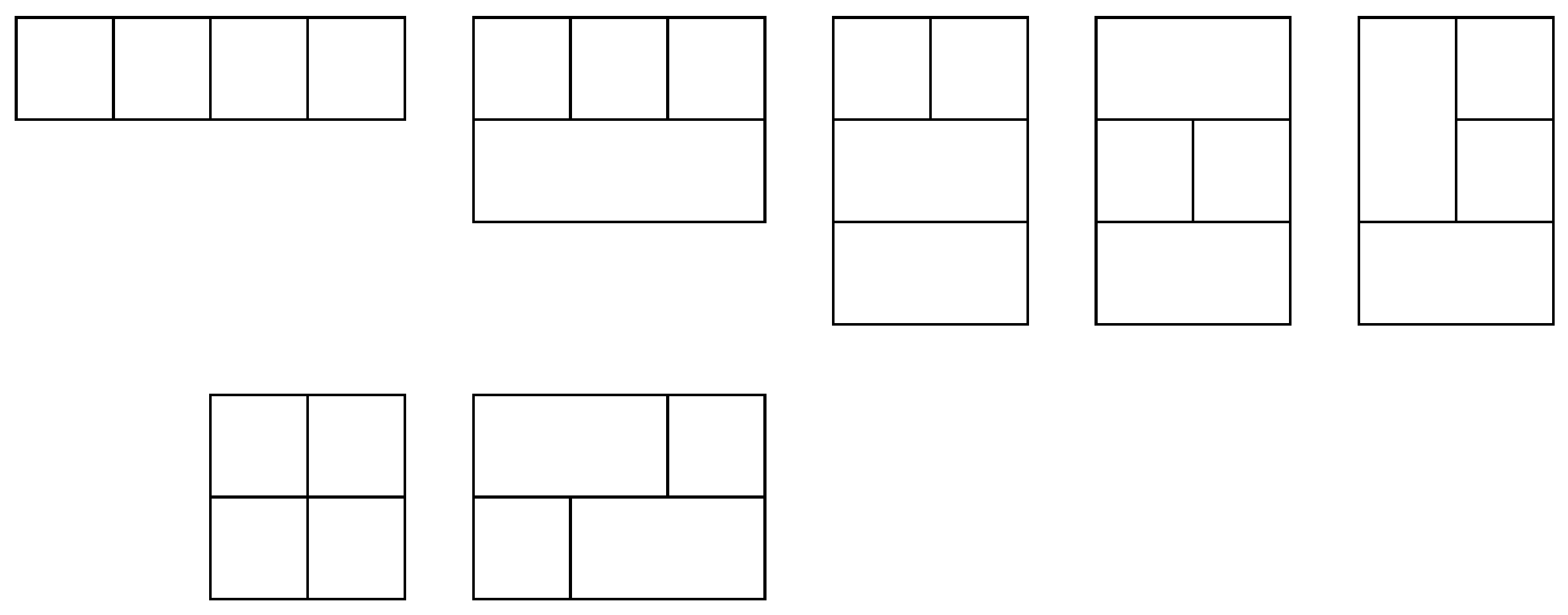}
\caption{All seven distinct Mondrian topologies at 4-loop.} \label{fig-9}
\end{center}
\end{figure}

In figure \ref{fig-9}, we list all seven distinct Mondrian topologies which include five (generalized) ladders,
one cross and one brick-wall. Once an ordered subspace is chosen, say $Y(1234)$ (fixing $X(1234)$), we then can
enumerate all possible orientations of these topologies with all loop numbers in their boxes fixed.
For each Mondrian topology in general there are eight orientations due to the \textit{dihedral symmetry} \cite{Dihedral}
which includes rotations by multiples of 90 degrees and left-right (or up-down) reflection,
while additional symmetries for some particular topologies can reduce this number to four, two or even one.

\begin{figure}
\begin{center}
\includegraphics[width=0.85\textwidth]{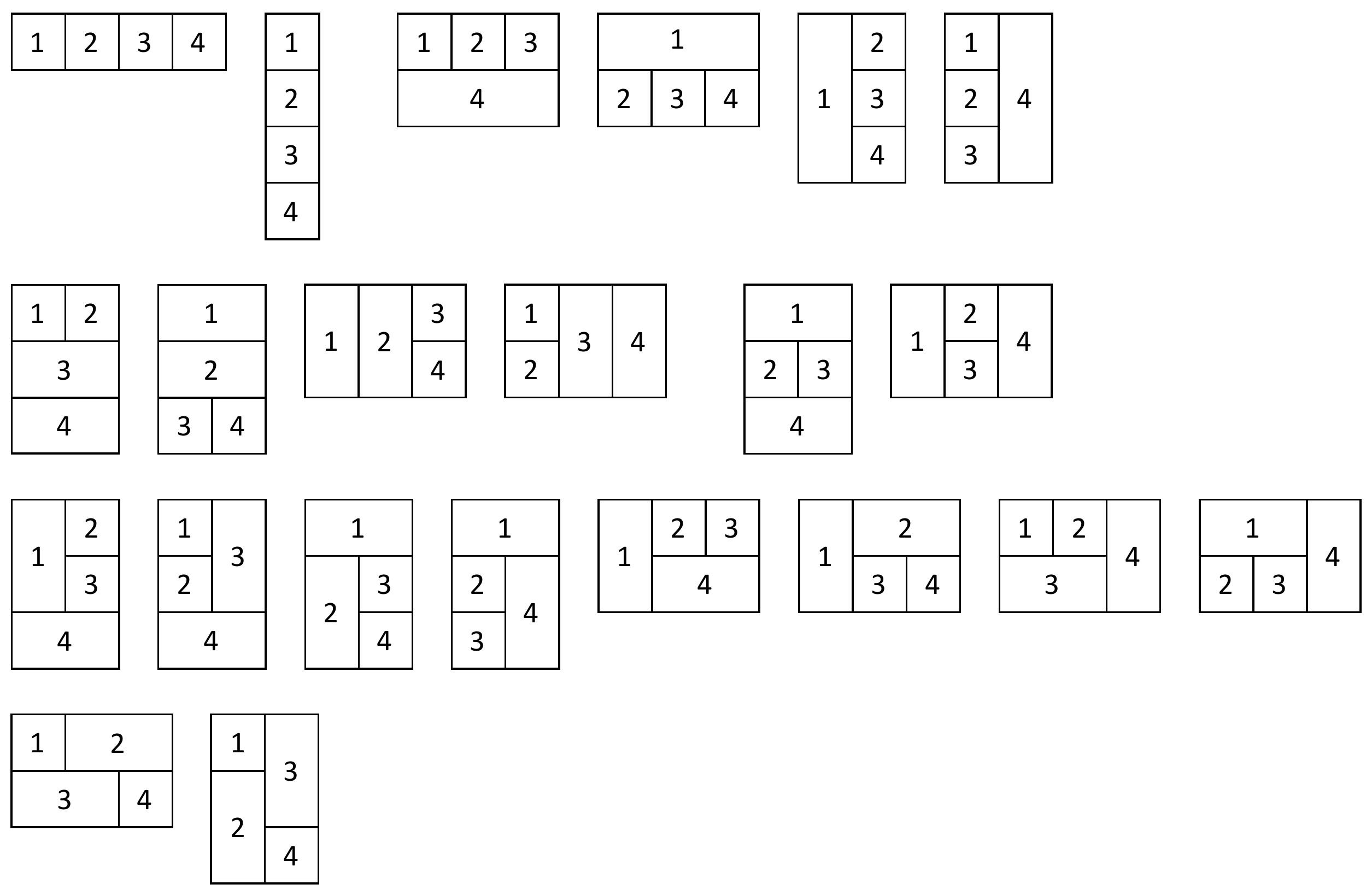}
\caption{All possible Mondrian diagrams at 4-loop in subspace $Y(1234)$.} \label{fig-10}
\end{center}
\end{figure}

\begin{figure}
\begin{center}
\includegraphics[width=0.37\textwidth]{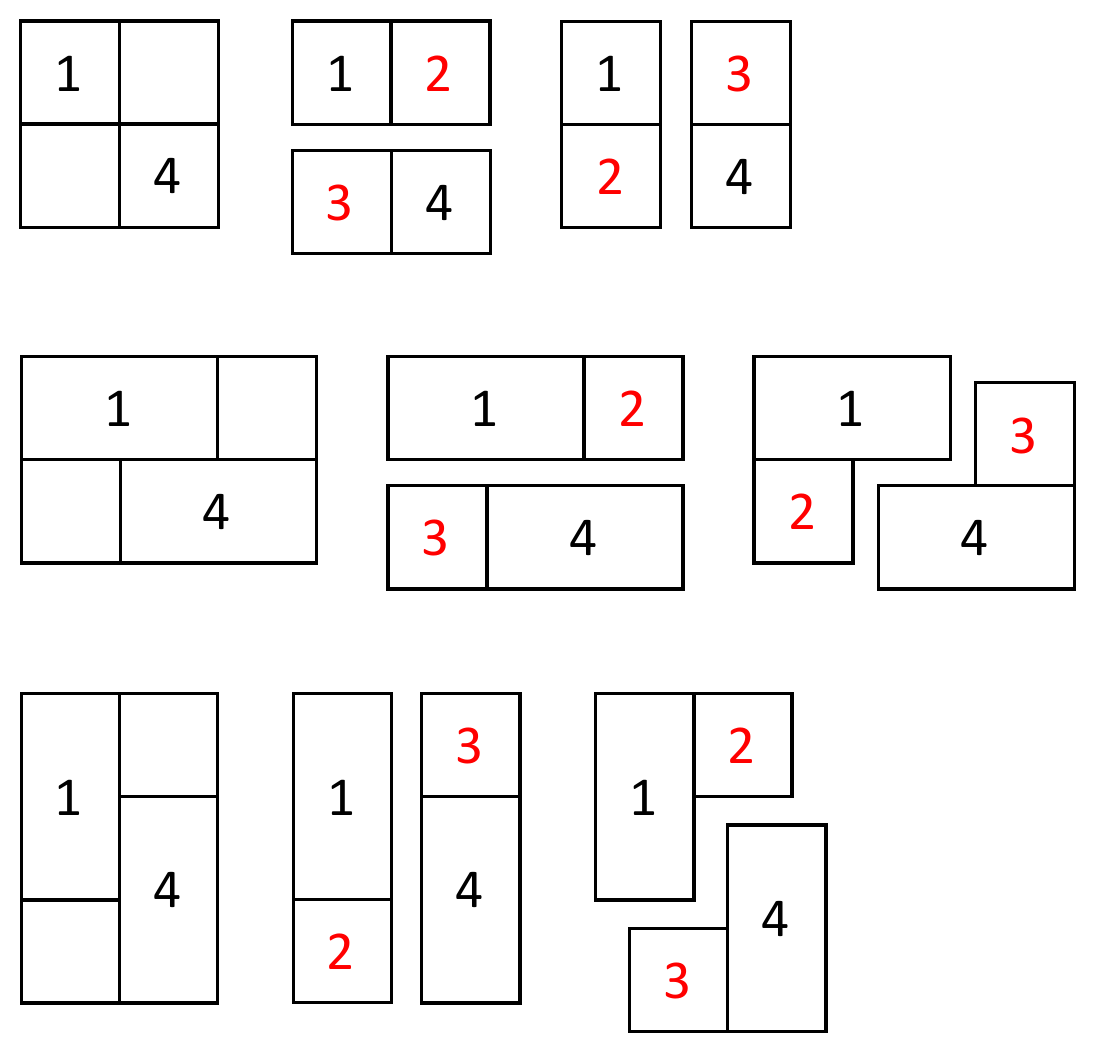}
\caption{Forbidden Mondrian diagrams in subspace $Y(1234)$.} \label{fig-11}
\end{center}
\end{figure}

In figure \ref{fig-10}, we list all possible Mondrian diagrams of six distinct topologies in subspace $Y(1234)$,
note the cross is forbidden along with two brick-walls, as shown in figure \ref{fig-11} which also demonstrates
the reason of their absence. It is worth emphasizing that, in the last row of figure \ref{fig-10} we also have
two brick-walls, but after a left-right (or up-down) reflection they become inconsistent with $Y(1234)$.
In the first row of figure \ref{fig-11}, we explain this via a contradiction of equivalent
reasonings. First of all, the positions of numbers $1,4$ are fixed.
If we chop the cross into two ``composite'' boxes, so that they have a vertical contact,
numbers $2,3$ must be put in the NE and SW corners. But if we let them have a horizontal contact instead,
$2,3$ must be put in the SW and NE corners. Therefore, this diagram cannot exist.
For the second and third rows of figure \ref{fig-11}, it is similar but a bit more tricky, since now
boxes $1,4$ have a vertical or horizontal contact. Still, $2,3$ have no contact so their relation can only be determined
through an intermediate box 1 or 4, and there is no consistent way to satisfy orderings
$X(1234)$ and $Y(1234)$ simultaneously. The identical orderings for
both directions enforce a NW-SE positioning of $2,3$, while the NE-SW positioning of
their corresponding boxes simply excludes the former. In contrast, in the two brick-walls in figure \ref{fig-10},
boxes $2,3$ have a contact which helps eliminate such an inconsistency!

To ``unify'' these diagrams filled with numbers, we can check the completeness relation
after summing their Mondrian factors given by rules \eqref{eq-1}, explicitly as
\be
\bal
&D_{12}D_{13}D_{14}D_{23}D_{24}D_{34}\\
=\,&~~~\,X_{12}X_{23}X_{34}D_{13}D_{14}D_{24}+Y_{12}Y_{23}Y_{34}D_{13}D_{14}D_{24}\\
&+X_{12}X_{23}Y_{14}Y_{24}Y_{34}D_{13}+X_{23}X_{34}Y_{12}Y_{13}Y_{14}D_{24}
+X_{12}X_{13}X_{14}Y_{23}Y_{34}D_{24}+X_{14}X_{24}X_{34}Y_{12}Y_{23}D_{13}\\
&+X_{12}Y_{13}Y_{23}Y_{34}D_{14}D_{24}+X_{34}Y_{12}Y_{23}Y_{24}D_{13}D_{14}
+X_{12}X_{23}X_{24}Y_{34}D_{13}D_{14}+X_{13}X_{23}X_{34}Y_{12}D_{14}D_{24}\\
&+X_{23}Y_{12}Y_{24}Y_{13}Y_{34}D_{14}+X_{12}X_{24}X_{13}X_{34}Y_{23}D_{14}\\
&+X_{12}X_{13}Y_{14}Y_{23}Y_{34}D_{24}+X_{13}X_{23}Y_{12}Y_{24}Y_{34}D_{14}
+X_{23}X_{24}Y_{12}Y_{13}Y_{34}D_{14}+X_{24}X_{34}Y_{12}Y_{23}Y_{14}D_{13}\\
&+X_{12}X_{23}X_{14}Y_{24}Y_{34}D_{13}+X_{12}X_{13}X_{34}Y_{23}Y_{24}D_{14}
+X_{12}X_{24}X_{34}Y_{13}Y_{23}D_{14}+X_{14}X_{23}X_{34}Y_{12}Y_{13}D_{24}\\
&+X_{12}X_{34}Y_{13}Y_{23}Y_{24}D_{14}+X_{13}X_{23}X_{24}Y_{12}Y_{34}D_{14}.
\eal
\ee
Indeed, it confirms that we have taken all possible Mondrian diagrams into account, in subspace $Y(1234)$.
But how can we use these quantities made of $X$, $Y$ and $D$ for more physical purposes or more concretely,
to integrate them with the amplituhedron? Again, we have to go back to the original definitions \eqref{eq-2},
or in general,
\be
X_{ij}=X_{ji}=(x_j-x_i)(z_i-z_j),~Y_{ij}=Y_{ji}=(y_j-y_i)(w_i-w_j),
\ee
note that even if we distinguish $X_{ij}$ from $X_{ji}$ in rules \eqref{eq-1}, in terms of $x$ and $z$ they are in
fact identical, but we will keep this notational difference to remind that they represent different
ordered subspaces.

\begin{figure}
\begin{center}
\includegraphics[width=0.3\textwidth]{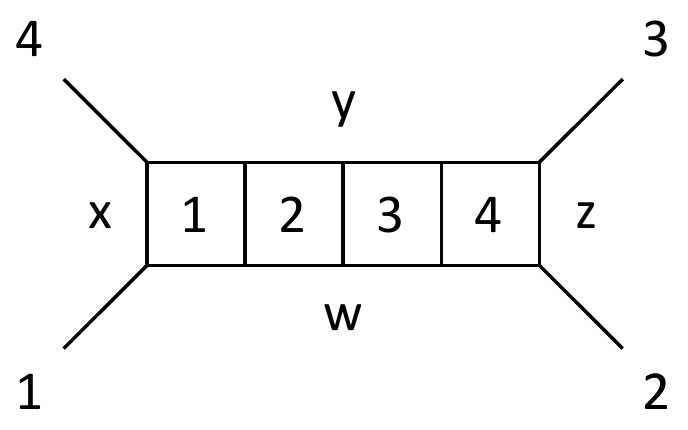}
\caption{A reminder of conventions for the true 4-loop ladder.} \label{fig-12}
\end{center}
\end{figure}

Now, we will analyze all six distinct Mondrian topologies one by one, by picking one example for each type
in figure \ref{fig-10}, and making extensive use of the properties of ordered subspaces, as done in the previous work
for the 3-loop amplituhedron. First, for the true 4-loop ladder, the 1st one in the 1st row of figure \ref{fig-10},
its Mondrian factor is $X_{12}X_{23}X_{34}D_{13}D_{14}D_{24}$. Obviously, this diagram represents all
ordered subspaces in which boxes $1,2,3,4$ are all parallel in the $y$ direction, so they are only ordered
in the $x$ direction. Recall the orderings of $z$ and $w$ are always opposite to those of $x$ and $y$ respectively,
then its corresponding $d\log$ form is
\be
X(1234)Z(4321)=\frac{1}{x_1x_{21}x_{32}x_{43}}\frac{1}{z_4z_{34}z_{23}z_{12}},
\ee
where $x_{ij}\!\equiv\!x_i\!-\!x_j$ and so forth,
and the form for $y$- and $w$-space is trivial since there is no ordering in the vertical
direction. Therefore, summing the Mondrian factor over the subspaces that admit it leads to
\be
x_{21}x_{32}x_{43}z_{34}z_{23}z_{12}\,D_{13}D_{14}D_{24}\times X(1234)Z(4321)
=D_{13}D_{14}D_{24}\,\frac{1}{x_1z_4},
\ee
and multiplying it by the rest physical poles, we get
\be
\bal
&D_{13}D_{14}D_{24}\,\frac{1}{x_1z_4}\times
\frac{1}{y_1y_2y_3y_4}\frac{1}{w_1w_2w_3w_4}\frac{1}{D_{12}D_{13}D_{14}D_{23}D_{24}D_{34}}\\
=\,&\frac{1}{x_1}\frac{1}{z_4}\frac{1}{y_1y_2y_3y_4}\frac{1}{w_1w_2w_3w_4}\frac{1}{D_{12}D_{23}D_{34}},
\eal
\ee
which is exactly the integrand of the true 4-loop ladder (not a generalized one).
Here, to use conventions consistent with the previous work, we remind the readers that
$x_i\!=\!\<A_iB_i\,14\>$, $y_i\!=\!\<A_iB_i\,34\>$, $z_i\!=\!\<A_iB_i\,23\>$, $w_i\!=\!\<A_iB_i\,12\>$
and $D_{ij}\!=\!\<A_iB_i\,A_jB_j\>$, as shown in figure \ref{fig-12}. Note the numbers in the outer rim
denote the external legs, while numbers inside the boxes, and $x,y,z,w$, denote faces surrounded
by external legs and internal propagators. It is trivial to see the denominator above matches all physical poles
of the diagram in figure \ref{fig-12}.

\begin{figure}
\begin{center}
\includegraphics[width=0.23\textwidth]{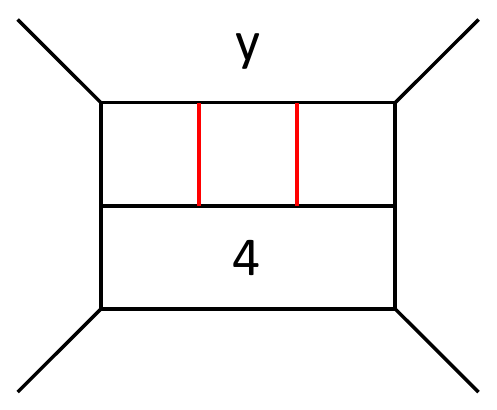}
\caption{Rung rule which counts extra connecting lines between zone 4 and $y$.} \label{fig-13}
\end{center}
\end{figure}

Then, for the next topology of ladders, say the 3rd one in the 1st row of figure \ref{fig-10},
its Mondrian factor is $X_{12}X_{23}Y_{14}Y_{24}Y_{34}D_{13}$. This diagram represents all
ordered subspaces in which boxes $1,2,3$ are parallel in the $y$ direction, while as a whole parallel with
box $4$ in the $x$ direction. Its corresponding form is
\be
X(123)Z(321)Y(\sg(123)\,4)W(4\,\sg(123))=\frac{1}{x_1x_{21}x_{32}}\frac{1}{z_3z_{23}z_{12}}
\frac{y_4^2}{y_1y_2y_3y_{41}y_{42}y_{43}}\frac{1}{w_4w_{14}w_{24}w_{34}},
\ee
where
\be
Y(\sg(123)\,4)=Y(1234)+Y(1324)+Y(2134)+Y(2314)+Y(3124)+Y(3214)=\frac{y_4^2}{y_1y_2y_3y_{41}y_{42}y_{43}}
\ee
can be straightforwardly checked, and it is analogous for
\be
W(4\,\sg(123))=W(4123)+W(4132)+W(4213)+W(4231)+W(4312)+W(4321)=\frac{1}{w_4w_{14}w_{24}w_{34}}.
\ee
Summing the Mondrian factor over the subspaces that admit it (including the trivial form for $x_4,z_4$),
and multiplying it by the rest physical poles leads to
\be
\bal
&X_{12}X_{23}Y_{14}Y_{24}Y_{34}D_{13}\times\frac{1}{x_4z_4}\,X(123)Z(321)Y(\sg(123)\,4)W(4\,\sg(123))
\times\frac{1}{D_{12}D_{13}D_{14}D_{23}D_{24}D_{34}}\\
=\,&\frac{1}{x_1x_4}\frac{1}{z_3z_4}\frac{y_4^2}{y_1y_2y_3}\frac{1}{w_4}\frac{1}{D_{12}D_{14}D_{23}D_{24}D_{34}},
\eal
\ee
again this is the correct integrand. Note the numerator $y_4^2$ is consistent with the
``rung rule'' in \cite{Bern:2005iz,Bern:2006ew,Bern:2007ct} as demonstrated in figure \ref{fig-13}.
In terms of Mondrian diagrams the rung rule has a clear pictorial meaning: it counts extra connecting lines between
two non-adjacent zones.

Pushing ahead for the next topology, say the 1st one in the 2nd row of figure \ref{fig-10},
this one in fact shows nothing new, so the following will be sketchy.
Summing its Mondrian factor $X_{12}Y_{13}Y_{23}Y_{34}D_{14}D_{24}$ over the subspaces that admit it
and multiplying it by the rest physical poles leads to
\be
\bal
&X_{12}Y_{13}Y_{23}Y_{34}D_{14}D_{24}\times\frac{1}{x_3x_4}\frac{1}{z_3z_4}\,X(12)Z(21)Y(\sg(12)\,34)W(43\,\sg(12))
\times\frac{1}{D_{12}D_{13}D_{14}D_{23}D_{24}D_{34}}\\
=\,&\frac{1}{x_1x_3x_4}\frac{1}{z_2z_3z_4}\frac{y_3}{y_1y_2}\frac{1}{w_4}\frac{1}{D_{12}D_{13}D_{23}D_{34}},
\eal
\ee
where we have used
\be
Y(\sg(12)\,34)=\frac{y_3}{y_1y_2y_{31}y_{32}y_{43}},~~W(43\,\sg(12))=\frac{1}{w_4w_{34}w_{13}w_{23}}.
\ee
Note the outstanding numerator $y_3$ again denotes the rung rule factor that represents an extra connecting line
between zone 3 and $y$.

\begin{figure}
\begin{center}
\includegraphics[width=0.18\textwidth]{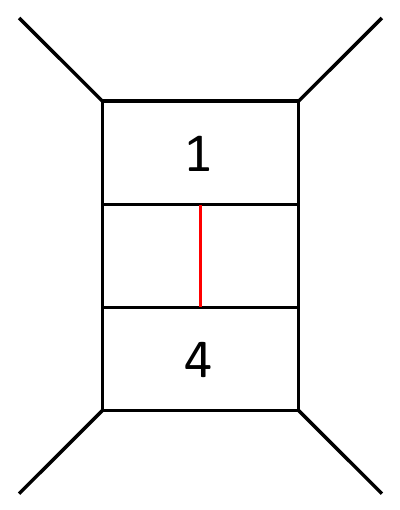}
\caption{The dual conformally invariant minimal replacement $y_{41}w_{14}\!\to\!D_{14}$ matches the rung rule.}
\label{fig-14}
\end{center}
\end{figure}

The next topology, say the 5th one in the 2nd row of figure \ref{fig-10}, is more tricky than the others.
Again, its corresponding sum is given by
\be
\bal
&X_{23}Y_{12}Y_{24}Y_{13}Y_{34}D_{14}\times\frac{1}{x_1x_4}\frac{1}{z_1z_4}\,X(23)Z(32)Y(1\,\sg(23)\,4)W(4\,\sg(23)\,1)
\times\frac{1}{D_{12}D_{13}D_{14}D_{23}D_{24}D_{34}}\\
=\,&\frac{1}{x_1x_2x_4}\frac{1}{z_1z_3z_4}\frac{y_{41}w_{14}}{y_1w_4}\frac{1}{D_{12}D_{13}D_{23}D_{24}D_{34}},
\eal
\ee
where we have used
\be
Y(1\,\sg(23)\,4)=\frac{y_{41}}{y_1y_{21}y_{31}y_{42}y_{43}},~~W(4\,\sg(23)\,1)=\frac{w_{14}}{w_4w_{24}w_{34}w_{12}w_{13}}.
\ee
There is a problem with the numerator $y_{41}w_{14}$: it is not dual conformally invariant.
A minimal treatment to heal this is to replace $y_{41}w_{14}$ by $D_{14}\!=\!(x_4\!-\!x_1)(z_1\!-\!z_4)\!+\!y_{41}w_{14}$.
If we set its explanation aside for the moment, the numerator $D_{14}$ happens to be the correct rung rule factor,
as demonstrated in figure \ref{fig-14}.

In general, if there are extra connecting lines between two loops (or internal faces), this substitution applies,
while for connecting lines between a loop and an external face that is surrounded by external legs,
there is no problem of dual conformal invariance as can be seen from previous examples.

\begin{figure}
\begin{center}
\includegraphics[width=0.19\textwidth]{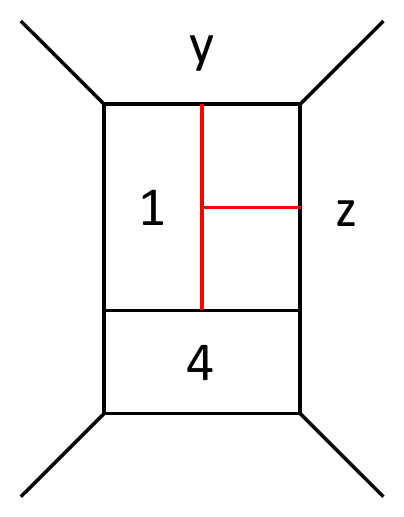}
\caption{Two rung rule factors $z_1$ and $y_4$.} \label{fig-15}
\end{center}
\end{figure}

For the next topology, say the 1st one in the 3rd row of figure \ref{fig-10}, its corresponding sum is given by
\be
\bal
&X_{12}X_{13}Y_{14}Y_{23}Y_{34}D_{24}\times\frac{1}{x_4z_4}\,X(1\,\sg(23))Z(\sg(23)\,1)
Y(\sg(1,23)\,4)W(4\,\sg(1,32))\times\frac{1}{D_{12}D_{13}D_{14}D_{23}D_{24}D_{34}}\\
=\,&\frac{1}{x_1x_4}\frac{z_1}{z_2z_3z_4}\frac{y_4}{y_1y_2}\frac{1}{w_4}\frac{1}{D_{12}D_{13}D_{14}D_{23}D_{34}},
\eal
\ee
where
\be
Y(\sg(1,23)\,4)=Y(1234)+Y(2134)+Y(2314)=\frac{y_4}{y_1y_2y_{32}y_{43}y_{41}}
\ee
represents the subspace in which $y_2\!<\!y_3$ and $y_1,y_2,y_3\!<\!y_4$, and it is analogous for
\be
W(4\,\sg(1,32))=W(4132)+W(4312)+W(4321)=\frac{1}{w_4w_{34}w_{14}w_{23}}.
\ee
Note that there are two rung rule factors for this topology, as indicated in figure \ref{fig-15}.

\begin{figure}
\begin{center}
\includegraphics[width=0.5\textwidth]{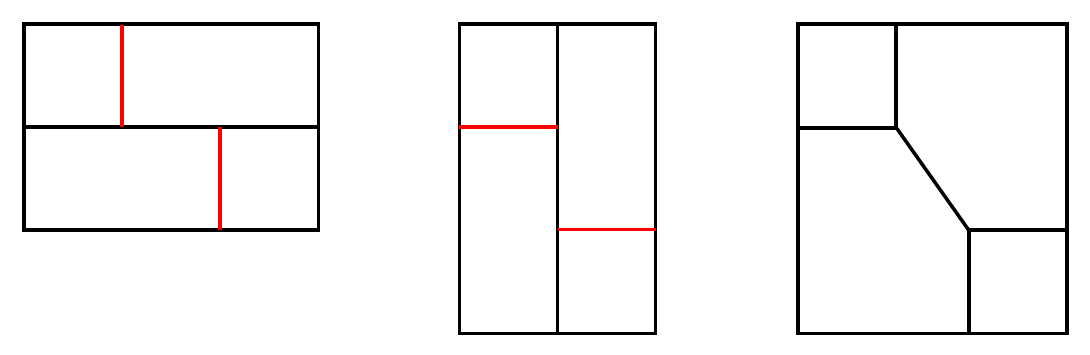}
\caption{Mondrian topologies automatically have a rung rule meaning.} \label{fig-16}
\end{center}
\end{figure}

For the last topology which is a brick-wall, say the 1st one in the 4th row of figure \ref{fig-10},
its corresponding sum is given by
\be
\bal
&X_{12}X_{34}Y_{13}Y_{23}Y_{24}D_{14}\times X(12)X(34)Z(21)Z(43)\,\frac{1}{y_{42}}\,Y(\sg(12)\,3)\,
\frac{1}{w_{13}}\,W(\sg(34)\,2)\times\frac{1}{D_{12}D_{13}D_{14}D_{23}D_{24}D_{34}}\\
=\,&\frac{1}{x_1x_3}\frac{1}{z_2z_4}\frac{y_3}{y_1y_2}\frac{w_2}{w_3w_4}\frac{1}{D_{12}D_{13}D_{23}D_{24}D_{34}}.
\eal
\ee
This topology also contains two rung rule factors, which are $y_3$ and $w_2$ for the particular diagram above.
From the brick-wall diagrams it is crucial to realize that, for Mondrian topologies whether an internal line is
horizontal or vertical makes a difference, since this distinguishes different rung rule factors. In figure \ref{fig-16},
we present three diagrams which are equivalent in the sense of \textit{ordinary} topology for Feynman diagrams,
but inequivalent as Mondrian topologies (or of inequivalent orientations of the same Mondrian topology), where
their central internal lines are horizontal, vertical and skew respectively.
While the third diagram has no Mondrian meaning, the first and second ones clearly indicate different rung rule factors.

This example shows how Mondrian diagrams refine ordinary Feynman diagrams, and integrate themselves with
dual conformal invariance which calls for the rung rule. In general, due to the existence of the rung rule factors
in local integrands, an ordinary diagram cannot fully reflect its numerator structure, but a Mondrian diagram can.
At 5-loop, we can find more such examples of various topologies \cite{Bern:2007ct}.

\begin{figure}
\begin{center}
\includegraphics[width=0.15\textwidth]{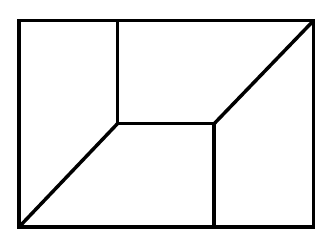}
\caption{The only non-Mondrian topology at 4-loop.} \label{fig-17}
\end{center}
\end{figure}

Now we have comprehensively understood the Mondrian diagrammatics in subspace $X(1234)Y(1234)$ at 4-loop.
There are two missing pieces still, one is the cross topology in figure \ref{fig-9} which will be discussed soon,
and the other is the non-Mondrian topology given in figure \ref{fig-17}
which like the cross is also associated with a minus sign \cite{Bern:2006ew}.
Though 4-vertices are allowed inside a Mondrian diagram, they are not allowed to appear on its rim as what
figure \ref{fig-17} shows, since there is no way to ``Mondrianize'' them at fixed external corners.

\begin{figure}
\begin{center}
\includegraphics[width=0.42\textwidth]{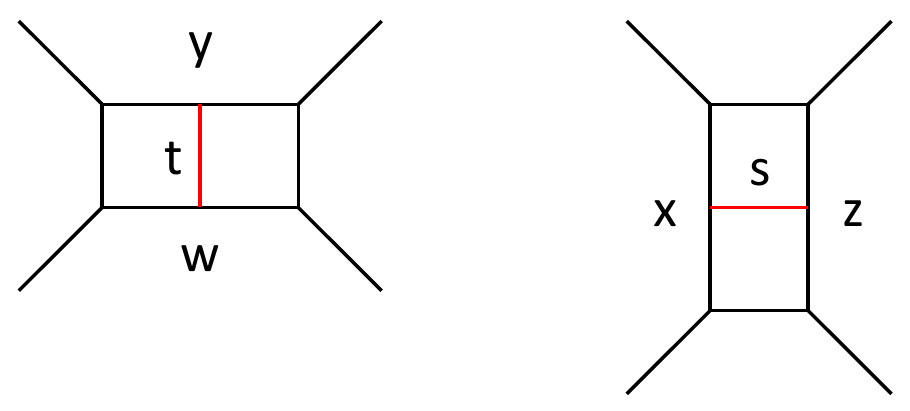}
\caption{Rung rule for connecting lines between two external faces.} \label{fig-18}
\end{center}
\end{figure}

Last, it is equally important to notice that, the rung rule also applies for connecting lines between two external faces
which will bring factors made of Mandelstam variables $s$ and $t$. In figure \ref{fig-18}, we present two simplest
2-loop examples to demonstrate this extra rule. If we use $s,t$ for the integrands, we also need the translation
between amplituhedron variables and the zone variables:
\be
x_i\to(q_{14}-q_i)^2,~y_i\to(q_{34}-q_i)^2,~z_i\to(q_{23}-q_i)^2,~w_i\to(q_{12}-q_i)^2,~D_{ij}\to(q_i-q_j)^2, \labell{eq-4}
\ee
where each $q$ is defined via a face surrounded by external legs and/or internal propagators. The difference
of two $q$'s is the propagator's momentum along the border of two corresponding adjacent faces.
Moreover, there is an overall factor of $st$ for all loop integrands. Collecting everything,
we now can map all Mondrian diagrams to familiar dual conformally invariant
integrands in \cite{Bern:2005iz,Bern:2006ew,Bern:2007ct}.

\subsection{Ordered subspace $X(1234)Y(1324)$}

\begin{figure}
\begin{center}
\includegraphics[width=0.65\textwidth]{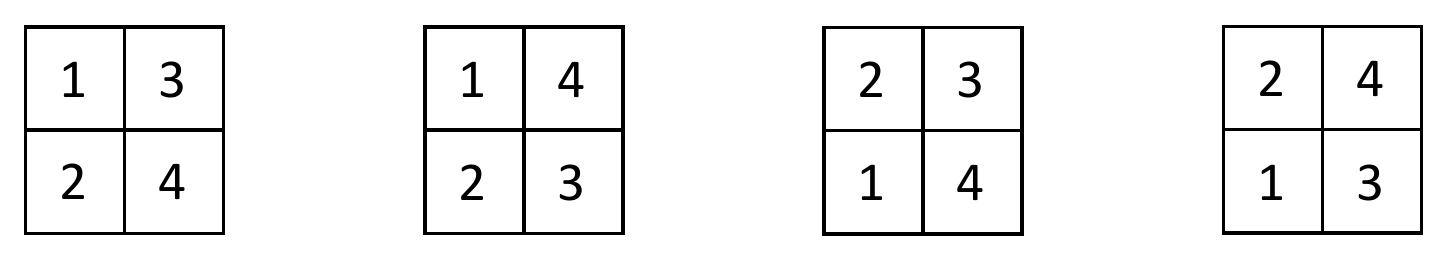}
\caption{Four possible combinations of numbers for the cross in $X(1234)$.} \label{fig-19}
\end{center}
\end{figure}

To understand the cross diagram's role in the completeness relation, we consider ordered subspace $Y(1324)$
for example. It is already known, identical orderings for both directions, namely
$X(\sg_1\sg_2\sg_3\sg_4)Y(\sg_1\sg_2\sg_3\sg_4)$ in general, exclude the cross. Now we reverse the problem to see,
what constraint the cross will impose on the relative ordering between $x$ and $y$, and show that $Y(1324)$ is
one of the admitted permutations.

If we keep fixing $X(1234)$ and chop the cross into two composite boxes which have a vertical contact,
numbers $1,2$ must stay in the left piece, and $3,4$ the right. Then there are four combinations as shown
in figure \ref{fig-19}, corresponding to subspaces
$Y(\sg(13)\,\sg(24))$, $Y(\sg(14)\,\sg(23))$, $Y(\sg(23)\,\sg(14))$ and $Y(\sg(24)\,\sg(13))$ respectively.
Therefore $Y(1324)$ is admitted while $Y(1234)$ is excluded. Once the cross is allowed to exist, the other
two brick-walls in figure \ref{fig-11} are also allowed, in each of which the pair of boxes without a contact has a
NE-SW positioning. With this subtlety clarified, we can continue the same check of the completeness relation.

\begin{figure}
\begin{center}
\includegraphics[width=0.84\textwidth]{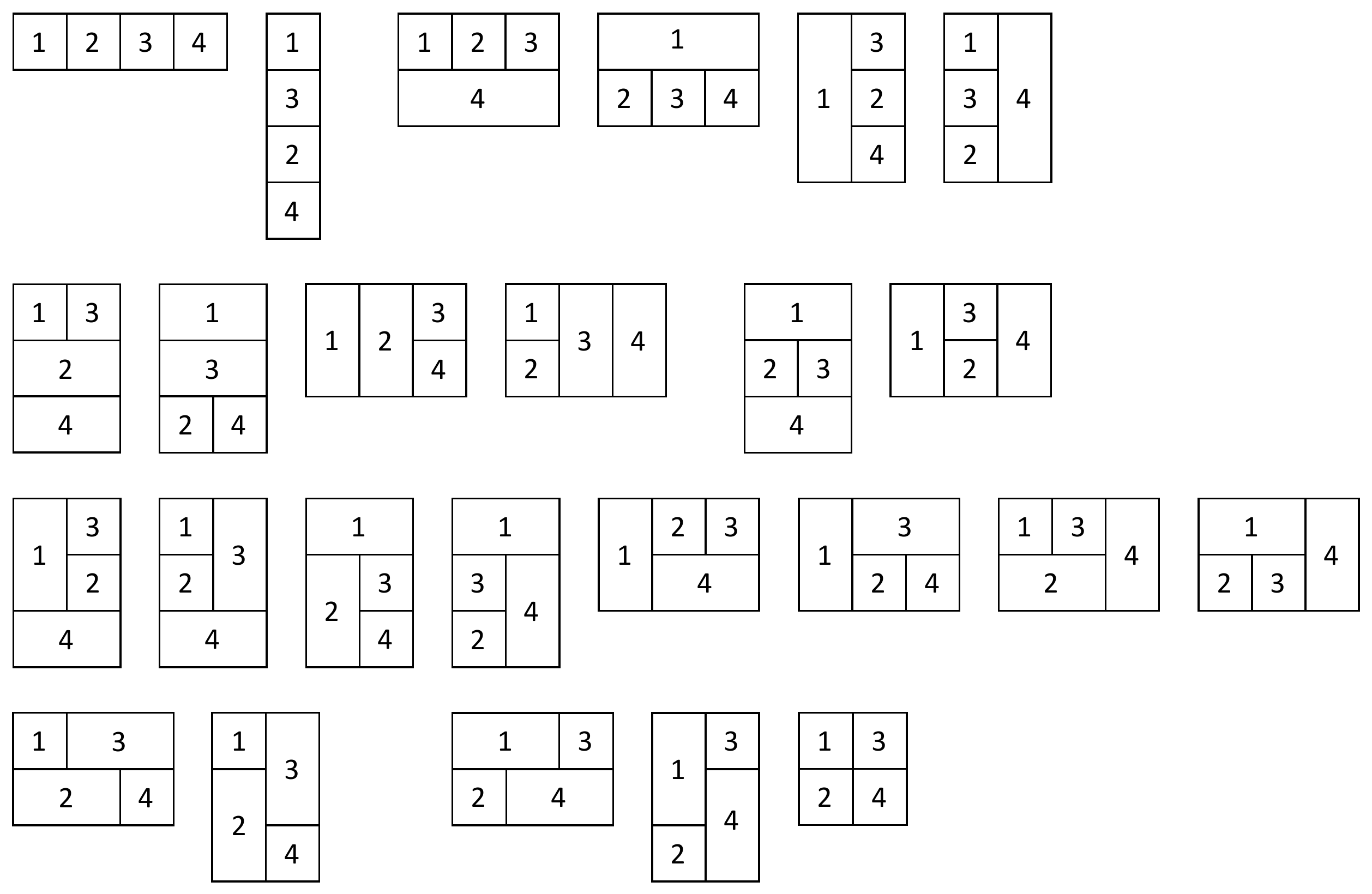}
\caption{All possible Mondrian diagrams at 4-loop in subspace $Y(1324)$.} \label{fig-20}
\end{center}
\end{figure}

In figure \ref{fig-20}, we list all possible Mondrian diagrams of seven distinct topologies including the cross,
in subspace $Y(1324)$, from which we confirm the following identity
\be
\bal
&D_{12}D_{13}D_{14}D_{23}D_{24}D_{34}\\
=\,&~~~\,X_{12}X_{23}X_{34}D_{13}D_{14}D_{24}+Y_{13}Y_{32}Y_{24}D_{12}D_{14}D_{34}\\
&+X_{12}X_{23}Y_{14}Y_{24}Y_{34}D_{13}+X_{23}X_{34}Y_{12}Y_{13}Y_{14}D_{24}
+X_{13}X_{12}X_{14}Y_{32}Y_{24}D_{34}+X_{14}X_{34}X_{24}Y_{13}Y_{32}D_{12}\\
&+X_{13}Y_{12}Y_{32}Y_{24}D_{14}D_{34}+X_{24}Y_{13}Y_{32}Y_{34}D_{12}D_{14}
+X_{12}X_{23}X_{24}Y_{34}D_{13}D_{14}+X_{13}X_{23}X_{34}Y_{12}D_{14}D_{24}\\
&+X_{23}Y_{12}Y_{24}Y_{13}Y_{34}D_{14}+X_{13}X_{34}X_{12}X_{24}Y_{32}D_{14}\\
&+X_{13}X_{12}Y_{14}Y_{32}Y_{24}D_{34}+X_{13}X_{23}Y_{12}Y_{24}Y_{34}D_{14}
+X_{23}X_{24}Y_{12}Y_{13}Y_{34}D_{14}+X_{34}X_{24}Y_{13}Y_{32}Y_{14}D_{12}\\
&+X_{12}X_{23}X_{14}Y_{24}Y_{34}D_{13}+X_{13}X_{12}X_{24}Y_{32}Y_{34}D_{14}
+X_{13}X_{34}X_{24}Y_{12}Y_{32}D_{14}+X_{14}X_{23}X_{34}Y_{12}Y_{13}D_{24}\\
&+X_{13}X_{24}Y_{12}Y_{32}Y_{34}D_{14}+X_{13}X_{23}X_{24}Y_{12}Y_{34}D_{14}+0,
\eal
\ee
where the last zero indicates a cancelation between two brick-walls and a cross as the latter is associated with
a minus sign, given by
\be
\bal
0&=X_{13}X_{24}Y_{12}Y_{14}Y_{34}D_{23}+X_{13}X_{14}X_{24}Y_{12}Y_{34}D_{23}-X_{13}X_{24}Y_{12}Y_{34}D_{14}D_{23}\\
&=X_{13}X_{24}Y_{12}Y_{34}D_{23}(Y_{14}+X_{14}-D_{14}),
\eal
\ee
as the key relation $D_{ij}\!=\!X_{ij}\!+\!Y_{ij}$ is used. There is nothing mysterious in adding a minus sign to
the cross so that we can use this relation to make it offset the two brick-walls, however, this trivial observation
can explain the minus sign of the corresponding physical integrand of a cross diagram! And more surprisingly, if we
associate each C-pattern with a minus sign, which means a Mondrian diagram with even number of C-patterns
has a plus sign, and odd number a minus sign, this also holds for its physical integrand up to 7-loop
as can be checked against \cite{Bern:2006ew,Bern:2007ct,Eden:2012tu,Bourjaily:2011hi}.

Analogous to the previous steps, it is easy to sum a cross diagram over all subspaces that admit it, so that we can obtain
its corresponding physical integrand. For the last diagram in figure \ref{fig-20}, we have
\be
\bal
&X_{13}X_{24}Y_{12}Y_{34}D_{14}D_{23}\times X(13)X(24)Z(31)Z(42)Y(12)Y(34)W(21)W(43)
\times\frac{1}{D_{12}D_{13}D_{14}D_{23}D_{24}D_{34}}\\
=\,&\frac{1}{x_1x_2}\frac{1}{z_3z_4}\frac{1}{y_1y_3}\frac{1}{w_2w_4}\frac{1}{D_{12}D_{13}D_{24}D_{34}}.
\eal
\ee
This completes the mapping between Mondrian diagrams and their corresponding physical integrands
for all seven distinct topologies at 4-loop, and it agrees with known results in \cite{Bern:2006ew}.

We have already discussed the existence condition of a cross (or C-pattern), along with its derivative brick-walls
(or B-patterns). But would there be a similar subtlety for the ladders?
The answer is no due to the crucial feature of ladders: we can
detach boxes from a ladder one by one while maintaining its exterior profile as a box at each step.
In other words, after each step of ``properly'' removing a box around its rim, the rest part is always
a legitimate sub-ladder, such that the filling of numbers can be proceeded until the sub-ladder has only
one single box left.

\begin{figure}
\begin{center}
\includegraphics[width=0.65\textwidth]{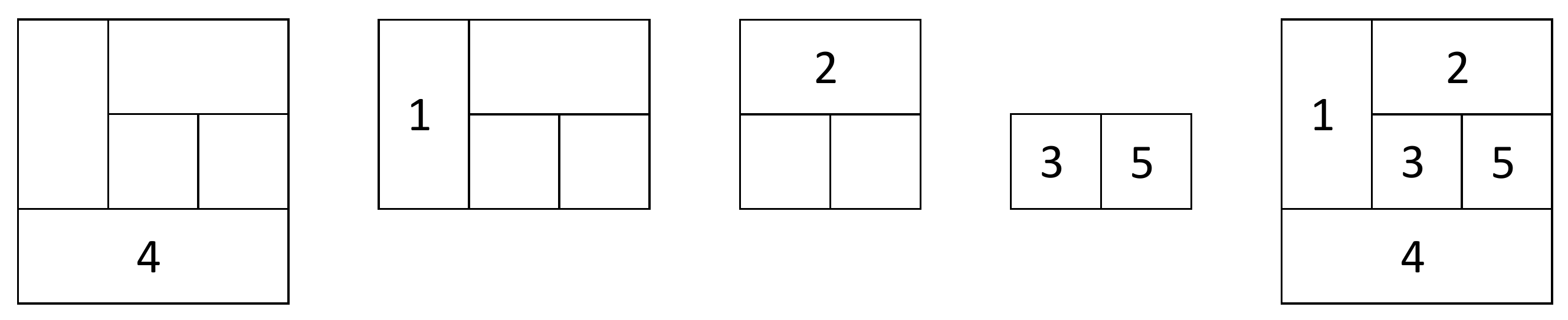}
\caption{Filling numbers step by step into a 5-loop ladder in subspace $Y(25314)$.} \label{fig-21}
\end{center}
\end{figure}

In figure \ref{fig-21}, we demonstrate this process by taking a previous example from figure \ref{fig-8},
a 5-loop ladder. Its subspace can be chosen to be $Y(25314)$ for example, and with $X(12345)$ we start this number filling:
\be
X(12345)Y(25314)\to X(1235)Y(2531)\to X(235)Y(253)\to X(35)Y(53),
\ee
where we delete numbers from both permutations after filling them into the boxes, one by one.
Therefore there is always one and only one configuration of the filled numbers for a ladder in general.

\subsection{Ordered subspace $X(1234)Y(2413)$ and its anomaly}

Besides the successful confirmation of completeness relations in $Y(1234)$ and $Y(1324)$, from which we also
comprehensively understand all seven Mondrian topologies at 4-loop, it is then appealing to further check
whether this neat identity holds for all ordered subspaces of $y$. In fact, instead of $4!\!=\!24$ permutations
we only need to check half of them, since $Y(\sg_1\sg_2\sg_3\sg_4)$ and its reverse $Y(\sg_4\sg_3\sg_2\sg_1)$
are identical if we reverse the $y$ direction while fixing $X(1234)$. Interestingly, we do find an exception,
namely the ordered subspace $Y(2413)$ (and its reverse $Y(3142)$).

\begin{figure}
\begin{center}
\includegraphics[width=0.84\textwidth]{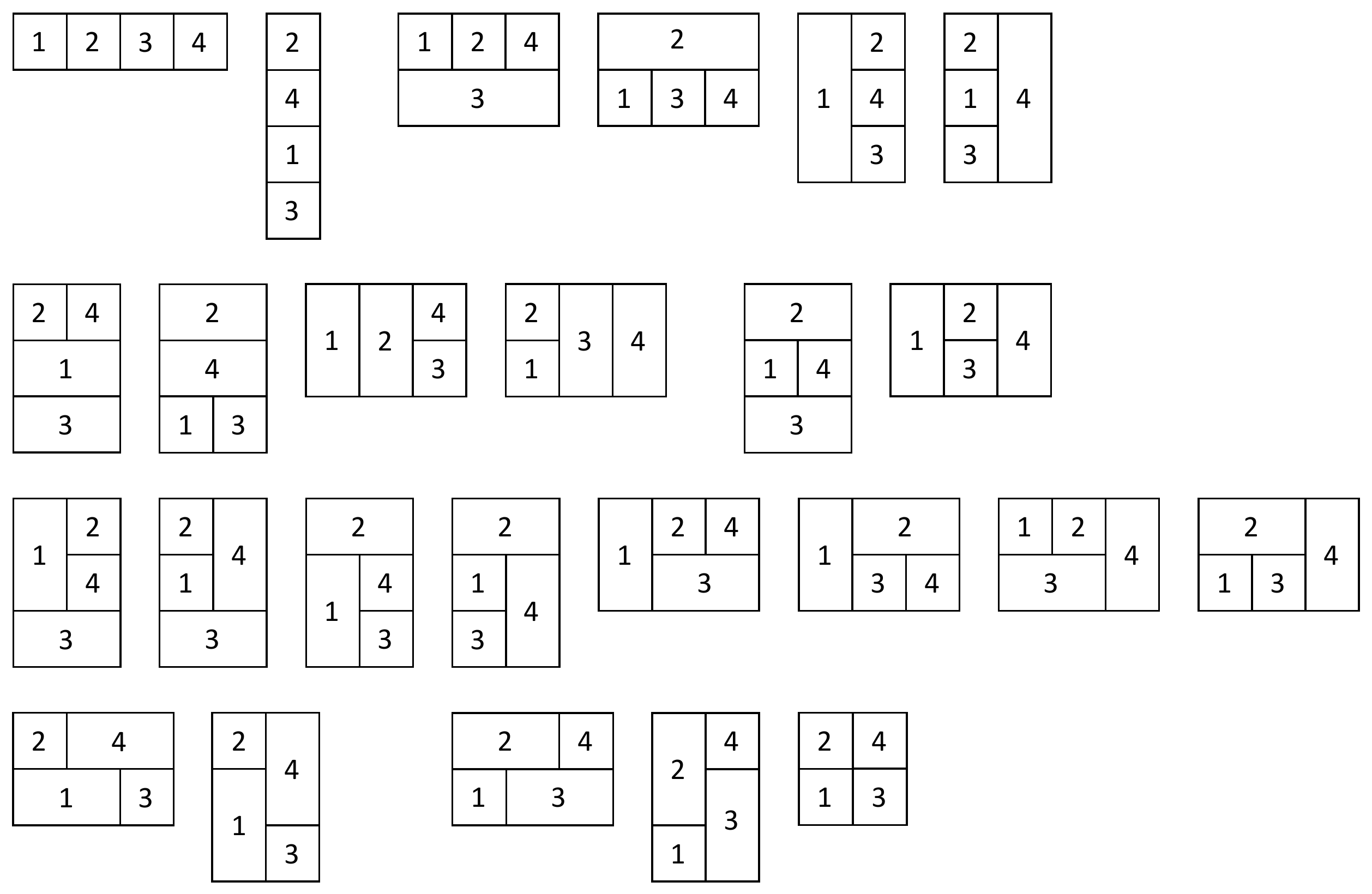}
\caption{All possible Mondrian diagrams at 4-loop in subspace $Y(2413)$.} \label{fig-22}
\end{center}
\end{figure}

In figure \ref{fig-22}, we list all possible Mondrian diagrams of seven distinct topologies
in subspace $Y(2413)$, in which the cross exists and offsets two brick-walls as usual.
But the completeness relation now fails as
\be
D_{12}D_{13}D_{14}D_{23}D_{24}D_{34}-\sum\textrm{(Mondrian factor)}
=X_{12}X_{34}Y_{24}Y_{13}(X_{14}Y_{23}-X_{23}Y_{41}) \labell{eq-5}
\ee
no longer vanishes. As we have explained in the introduction, this is due to the non-separable permutation $(2413)$
of $(1234)$:  there is no way to chop it into two
sub-permutations such that any number in one subset is always
larger (or smaller) than any one in the other. For $(2413)$, we have $(2)(413)$, $(24)(13)$ and $(241)(3)$ but none of
these partitions is separable according to the definition above. To understand further aspects,
we will return to this point when proceeding to the all-loop recursive proof of the completeness relation.

This type of non-zero results named as anomalies occur more frequently at 5-loop as we will soon see,
since the increasing cardinality admits more combinations that can form non-separable permutations.
The concept of separable permutations in mathematics is closely related to Schr\"{o}der numbers \cite{Separable},
which have a Mondrian diagrammatic interpretation as well. But this interpretation does not include C- or S-pattern
and there is no completeness relation either. Only with all four basic patterns and the relation
$D\!=\!X\!+\!Y$ inspired by amplituhedron, we may judge whether a permutation is separable or not, by its
completeness relation or anomaly otherwise.

\newpage
\section{Mondrian Diagrammatics at 5-loop}
\label{sec3}

We continue the same investigation of Mondrian diagrammatics at 5-loop again in two ordered subspaces $Y(12345)$
and $Y(14325)$ to present more concrete examples of the cancelation between crosses and brick-walls.
At 5-loop, the new spiral pattern starts to show up,
and there are now 15 distinct anomalies (or non-separable permutations) out of $5!/2\!=\!60$ combinations
modulo the trivial reversion in the $y$ direction. Note the spiral pattern or its simplest form as a pinwheel
topology, needs an extra treatment to correct it into a dual conformally invariant \textit{integral},
otherwise its invariance holds only at the integrand level.

\begin{figure}
\begin{center}
\includegraphics[width=0.6\textwidth]{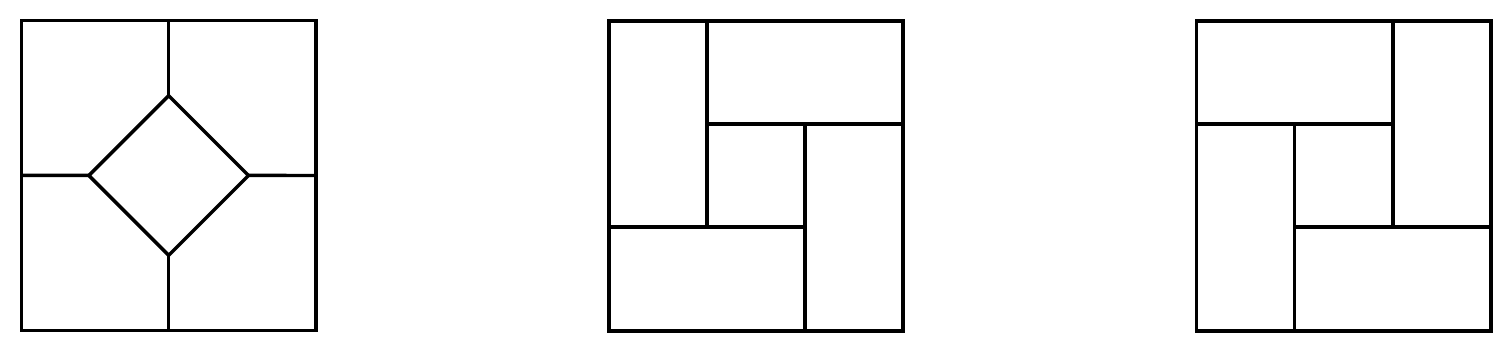}
\caption{Three forms of the pinwheel topology.} \label{fig-23}
\end{center}
\end{figure}

Pictorially, the pinwheel is in fact the Mondrianized form of a more symmetric form, given
in the left of figure \ref{fig-23}. It can be oriented either clockwise or counterclockwise, as shown in
the center and the right of figure \ref{fig-23} respectively. To Mondrianize this topology can help manifest
its rung rule, although this rung rule is \textit{different} from the one in \cite{Bern:2007ct} and that's why
we need the extra treatment mentioned above.

\subsection{Ordered subspaces $X(12345)Y(12345)$ and $X(12345)Y(14325)$}

\begin{figure}
\begin{center}
\includegraphics[width=0.95\textwidth]{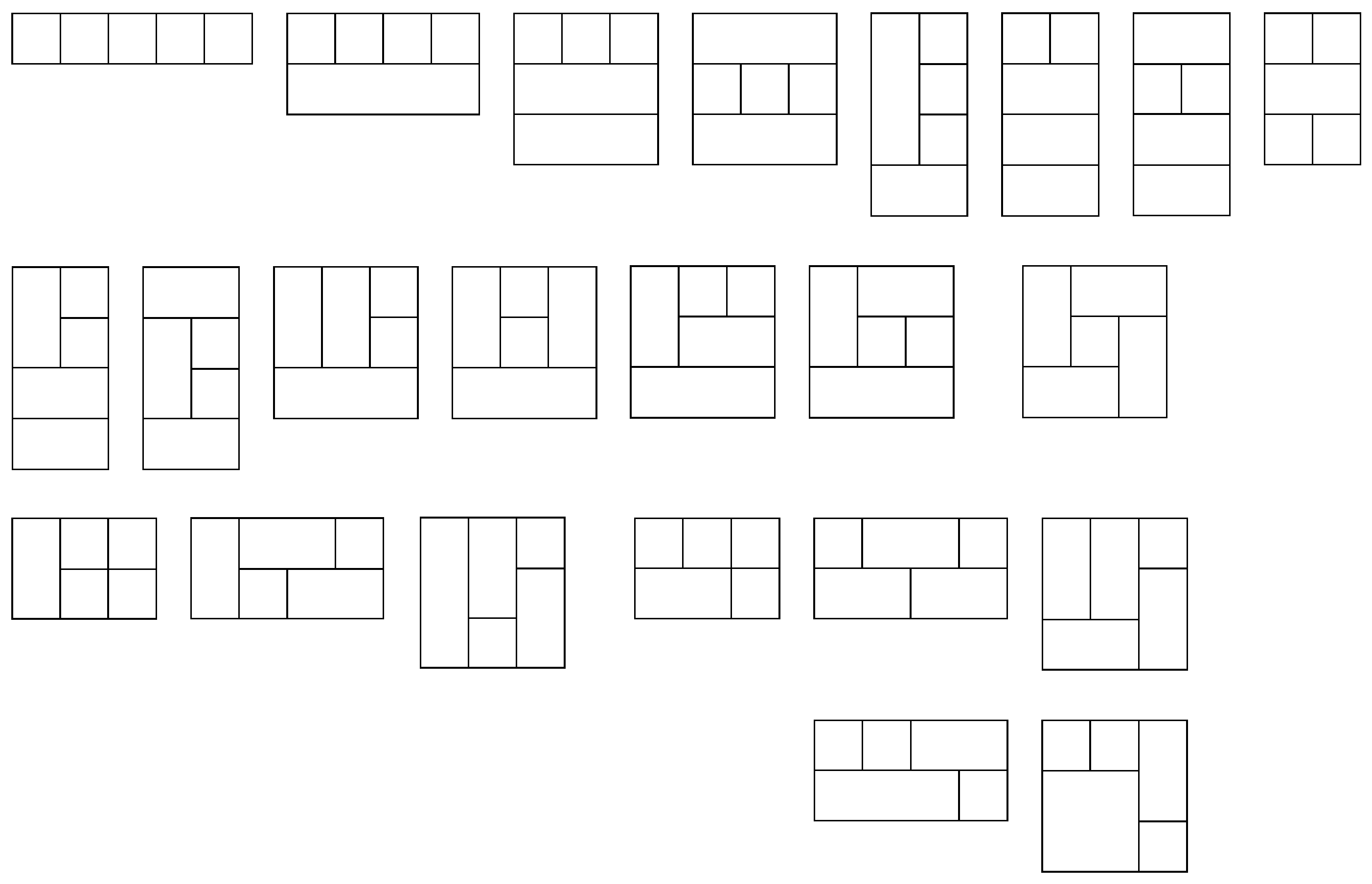}
\caption{All 23 distinct Mondrian topologies at 5-loop.} \label{fig-24}
\end{center}
\end{figure}

\begin{figure}
\begin{center}
\includegraphics[width=0.99\textwidth]{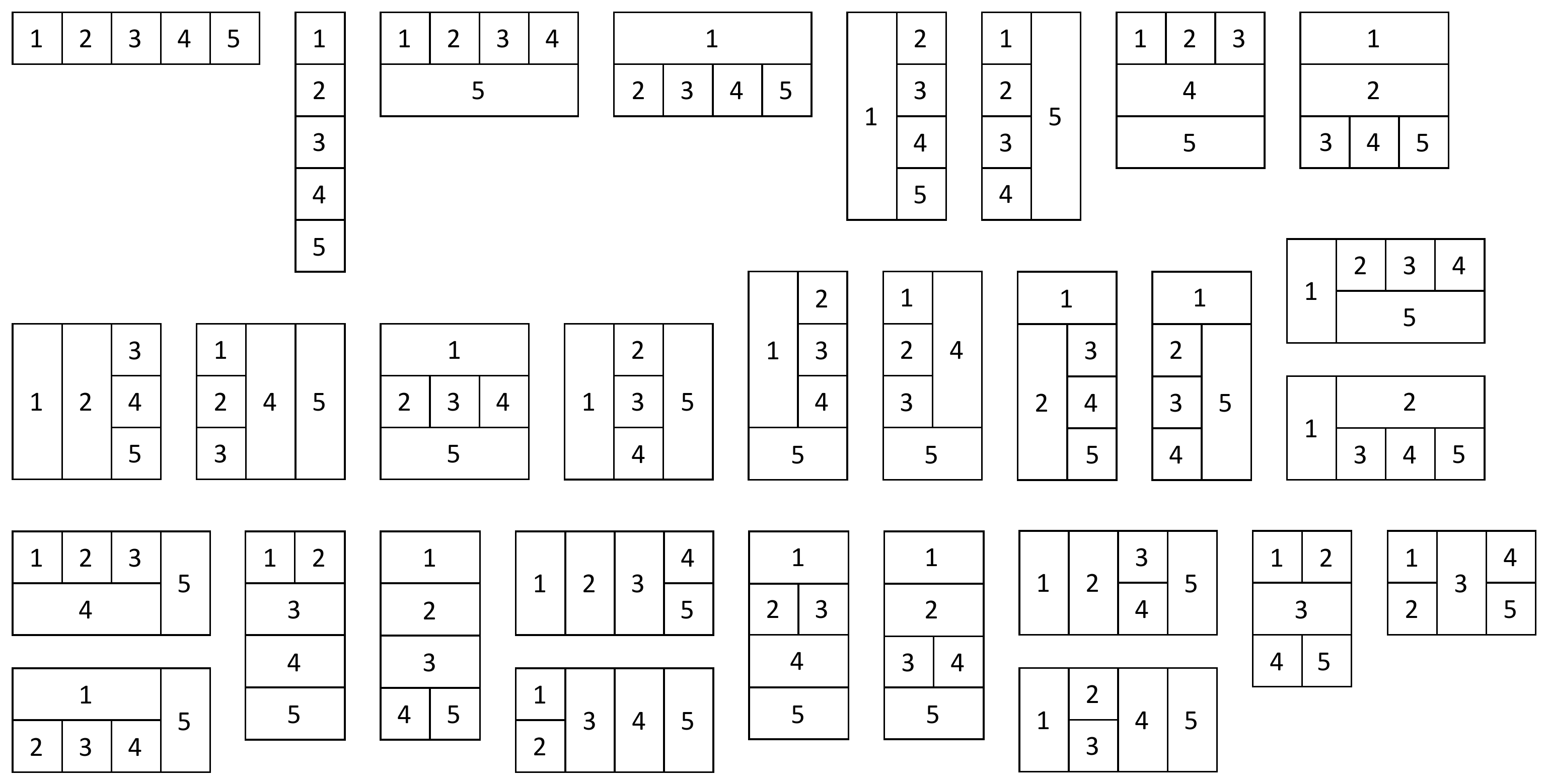}
\caption{All possible Mondrian diagrams at 5-loop in subspace $Y(12345)$: part 1/3.} \label{fig-25}
\end{center}
\end{figure}

\begin{figure}
\begin{center}
\includegraphics[width=0.97\textwidth]{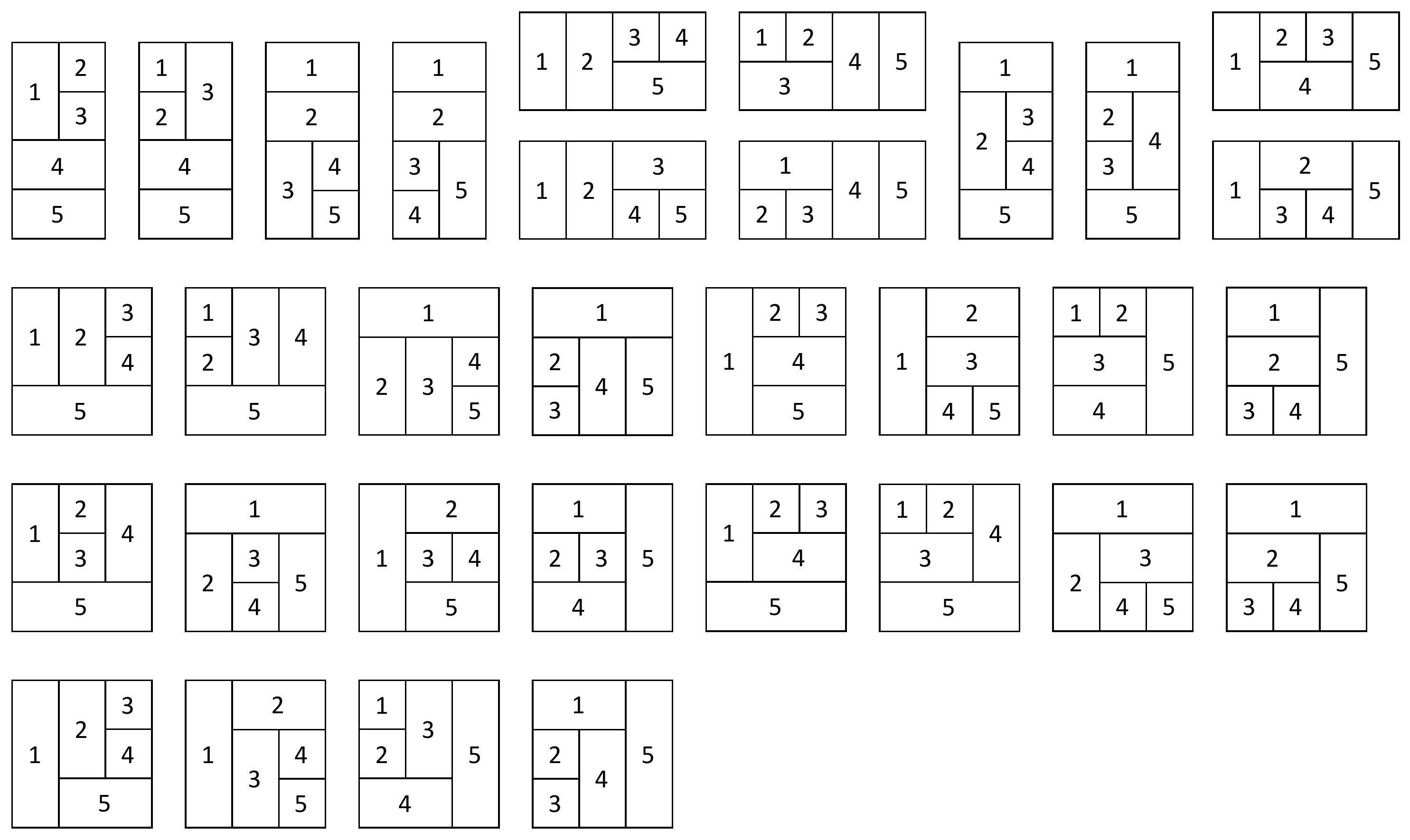}
\caption{All possible Mondrian diagrams at 5-loop in subspace $Y(12345)$: part 2/3.} \label{fig-26}
\end{center}
\end{figure}

\begin{figure}
\begin{center}
\includegraphics[width=1.04\textwidth]{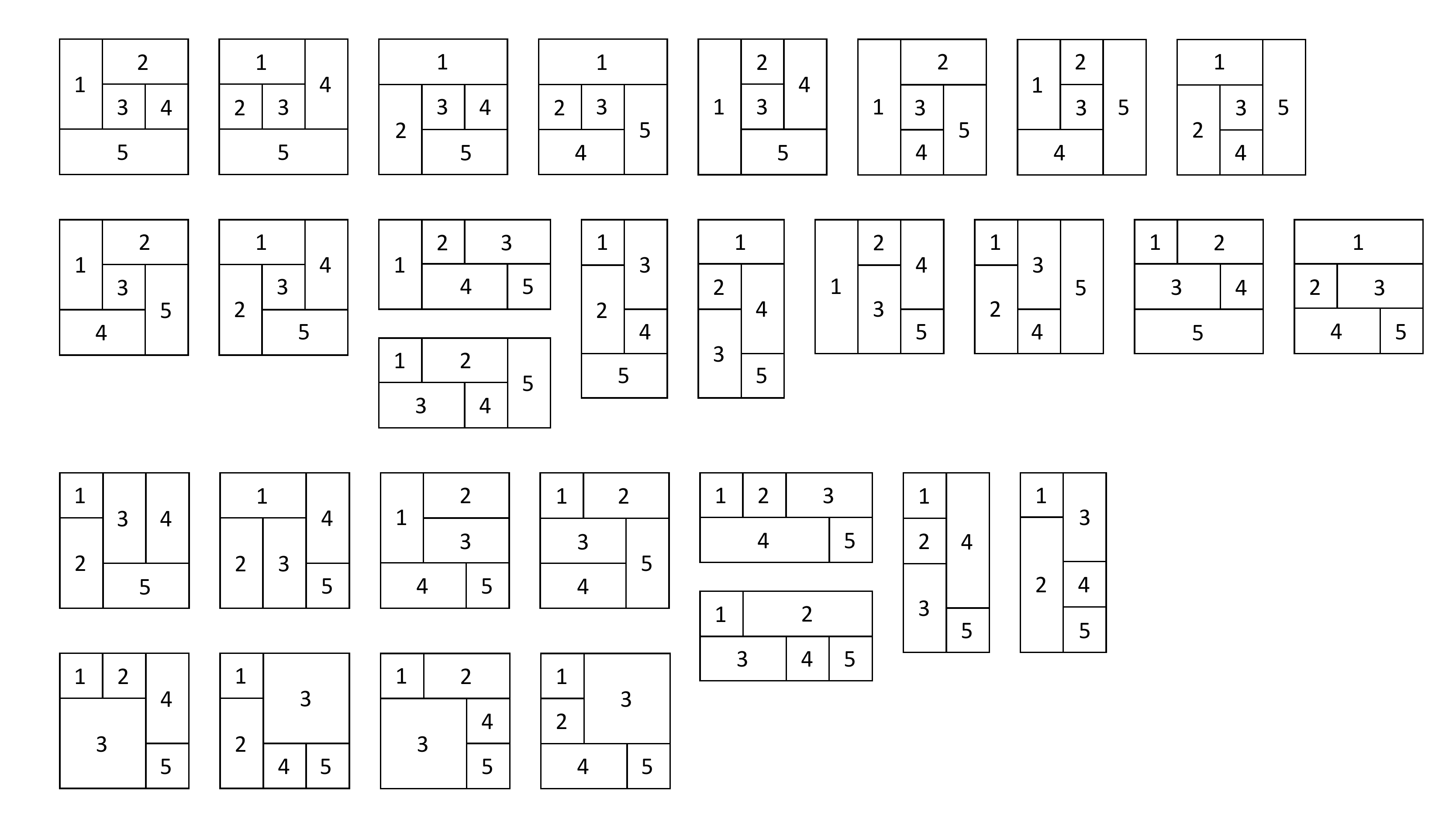}
\caption{All possible Mondrian diagrams at 5-loop in subspace $Y(12345)$: part 3/3.} \label{fig-27}
\end{center}
\end{figure}

Again, we start with the simplest ordered subspace $Y(12345)$ at 5-loop.
In figure \ref{fig-24}, we list all 23 distinct Mondrian topologies which include 14 ladders, 1 spiral,
2 crosses and 6 derivative brick-walls. In $Y(12345)$, by dihedral symmetry we can enumerate
all possible orientations of them filled with loop numbers as given in figures
\ref{fig-25}, \ref{fig-26} and \ref{fig-27}. From these objects we confirm the identity
\be
D_{12}D_{13}D_{14}D_{15}D_{23}D_{24}D_{25}D_{34}D_{35}D_{45}-\sum\textrm{(Mondrian factor)}=0.
\ee
The identical orderings for both directions again exclude the brick-walls
in each of which the pair of boxes without a contact has a NE-SW positioning. And in particular
the 5th topology in the 3rd row of figure \ref{fig-24} which is a brick-wall, has been totally excluded,
since no matter how it is oriented there is always one pair of non-contacting boxes that has a NE-SW positioning.
Note that for B-patterns, \textit{non-contacting} precisely means the pair of boxes
cannot be connected by a chain of boxes that align either horizontally or vertically.

\begin{figure}
\begin{center}
\includegraphics[width=1.06\textwidth]{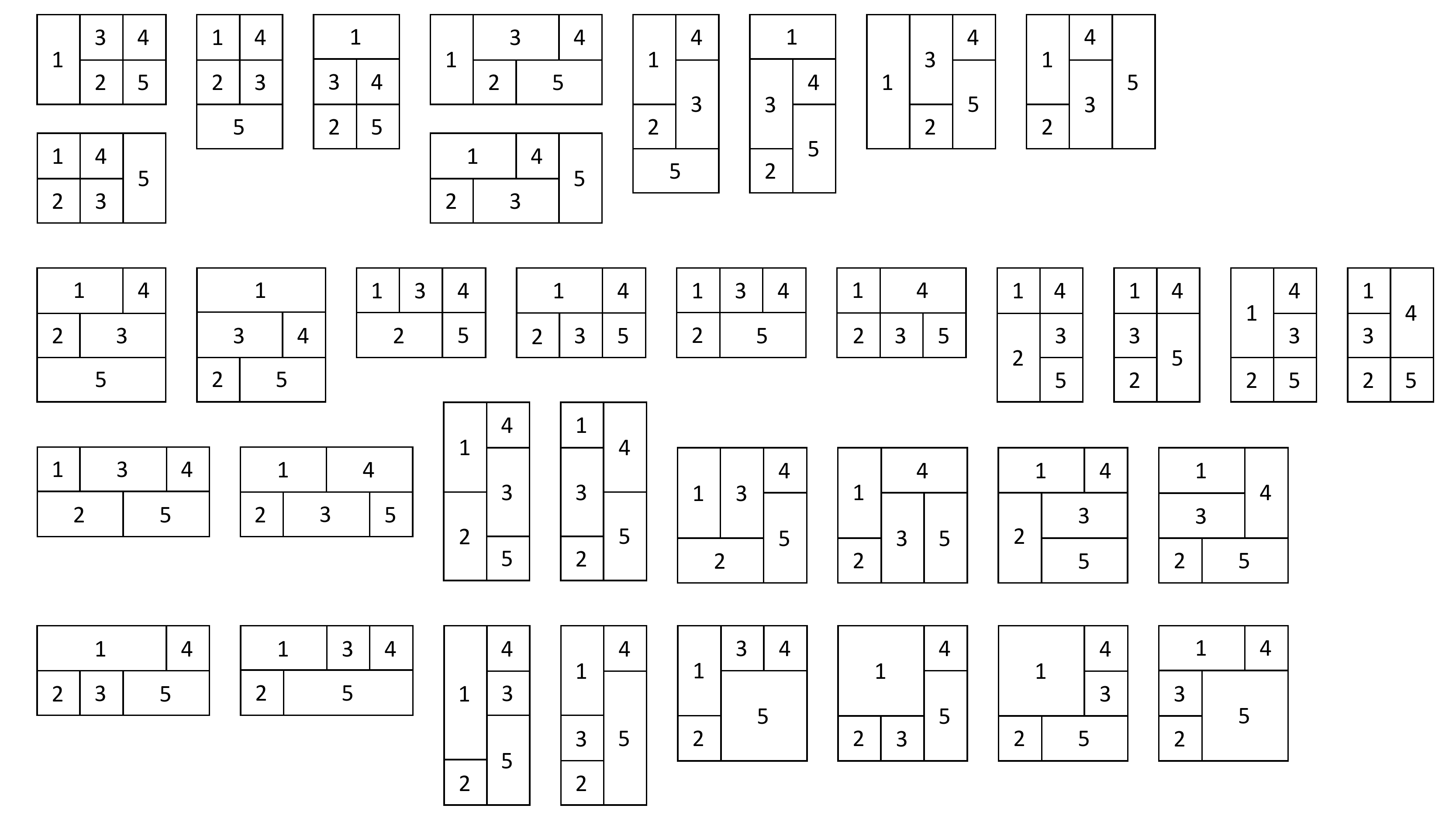}
\caption{Additional Mondrian diagrams that neatly cancel at 5-loop in subspace $Y(14325)$.} \label{fig-28}
\end{center}
\end{figure}

To see the presence of all Mondrian topologies we can switch to ordered subspace $Y(14325)$, analogous to
what we have done from $Y(1234)$ to $Y(1324)$ at 4-loop. The additional Mondrian diagrams that exist
in $Y(14325)$ are given in figure \ref{fig-28}. With the minus signs for the crosses,
all these contributions neatly cancel so that they will not affect the completeness relation.
For diagrams already exist in $Y(12345)$, it is trivial to rearrange the numbers in relevant boxes
to maintain the completeness relation.

Again, at 5-loop there are some topologies that have extra connecting lines between two loops and we need the
minimal treatment introduced for the 4-loop case. For example, there is a factor $(y_{51}w_{15})^2$ after we
complete the summation for the 3rd diagram in the 2nd row of figure \ref{fig-25}, which results from
\be
Y(1\,\sg(234)\,5)=\frac{y_{51}^2}{y_1y_{21}y_{31}y_{41}y_{52}y_{53}y_{54}},~~
W(5\,\sg(234)\,1)=\frac{w_{15}^2}{w_5w_{25}w_{35}w_{45}w_{12}w_{13}w_{14}}.
\ee
Then to heal its dual conformal invariance, we can trivially replace $(y_{51}w_{15})^2$ by $D_{15}^2$, which is
the correct rung rule factor.

\begin{figure}
\begin{center}
\includegraphics[width=0.37\textwidth]{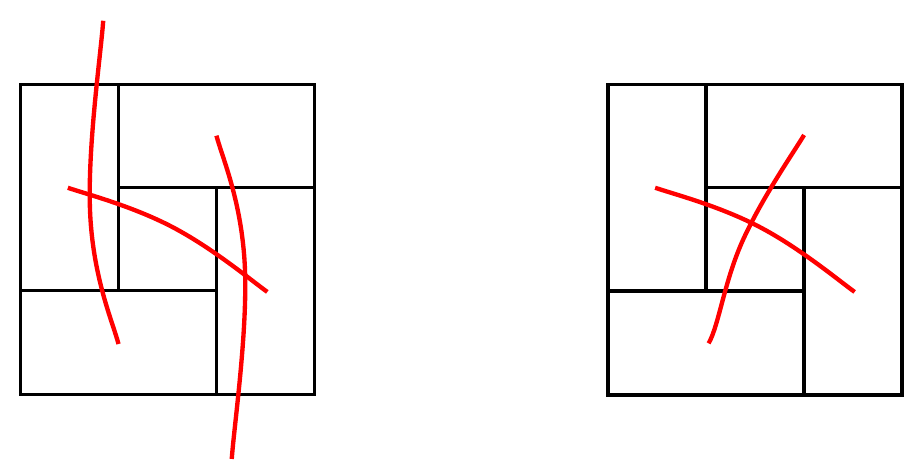}
\caption{Diagrammatic representations of two different integrands sharing the pinwheel's pole structure.} \label{fig-29}
\end{center}
\end{figure}

A new pattern at 5-loop is the spiral topology, and it also needs a special treatment required by dual conformal invariance.
For example, the summation for the 1st diagram in the 2nd row of figure \ref{fig-27} is given by
(it is equivalent to the rung rule in our context, but different from that in \cite{Bern:2007ct})
\be
\bal
&X_{12}X_{13}X_{35}X_{45}Y_{14}Y_{23}Y_{34}Y_{25}D_{15}D_{24}
\times\frac{x_5}{x_{21}x_4x_{54}}\,X(135)\,\frac{z_1}{z_{45}z_2z_{12}}\,Z(531)\,
\frac{y_4}{y_{52}y_1y_{41}}\,Y(234)\,\frac{w_2}{w_{14}w_5w_{25}}\,W(432)\\
&\times\frac{1}{D_{12}D_{13}D_{14}D_{15}D_{23}D_{24}D_{25}D_{34}D_{35}D_{45}}
=\frac{x_5}{x_1x_4}\frac{z_1}{z_2z_5}\frac{y_4}{y_1y_2}\frac{w_2}{w_4w_5}
\frac{1}{D_{12}D_{13}D_{14}D_{23}D_{25}D_{34}D_{35}D_{45}}, \labell{eq-3}
\eal
\ee
which is a dual conformally invariant integrand, however, its integral is not! Under certain off-shell limits
\cite{Bourjaily:2011hi,Drummond:2007aua,Nguyen:2007ya}, it is divergent. To see this, let zone variables
$q_1,q_2,q_4,q_5$ approach $q_3$ in the example above, then we encounter a logarithmic divergence
\be
\int\frac{\rho^{4\times4-1}\,d\rho}{\rho^{2\times8}}=\int\frac{d\rho}{\rho}\,~\textrm{as}~\rho\to0,
\ee
where $\rho$ is the radial coordinate characterizing the proximity between $q_1,q_2,q_4,q_5$ and $q_3$,
and the denominator results from the eight $D_{ij}$'s that approach zero. To heal this divergence, there are three choices
of minimal treatments for \eqref{eq-3}, as we can manipulate the numerator to offset it:
\be
x_5z_1\to D_{15},~\textrm{or}~y_4w_2\to D_{24},~\textrm{or both}~x_5z_1\to D_{15}~\textrm{and}~y_4w_2\to D_{24},
\labell{eq-10}
\ee
which provide an additional factor $\rho^2$ or $\rho^4$ while maintaining dual conformal invariance
at the integrand level. In terms of zone variables defined in \eqref{eq-4}, or in general,
$x_iz_j\!\to\!D_{ij}$ and $y_iw_j\!\to\!D_{ij}$ become
\be
(q_{14}-q_i)^2(q_{23}-q_j)^2\to s\,(q_i-q_j)^2,~~(q_{34}-q_i)^2(q_{12}-q_j)^2\to t\,(q_i-q_j)^2, \labell{eq-11}
\ee
so that the correct dimensions and scaling properties of dual conformal invariance are ensured.
While the first two choices have no essential difference, though they break the symmetry of the pinwheel in
different ways, the third choice is obviously different from the former two and it again preserves the symmetry.
We demonstrate this distinction in figure \ref{fig-29}, where the red curves indicate their numerators in terms of
zone variables (it is the same notation used in \cite{Nguyen:2007ya}, for example).

The only one unexplained fact is, the diagram in the right of figure \ref{fig-29} is associated with a minus sign.
This is known as the ``substitution rule'' in \cite{Bern:2007ct}, by which the cross diagram's minus sign at 4-loop
accounts for this one, as we obtain a pinwheel by inserting a box into its central 4-vertex (see the left
of figure \ref{fig-23}).

As we have remarked in the 4-loop case, Mondrian topologies automatically have a rung rule meaning
and now there are more examples at 5-loop. The 6 derivative brick-walls in the 3rd and 4th rows of figure \ref{fig-24}
correspond to $I_{17},I_{18},I_{11},I_{12},I_{14},I_{13}$ in \cite{Bern:2007ct} respectively
where $I_{17},I_{18}$ share the same ordinary topology for Feynman diagrams but with different rung rule factors,
so do $I_{11},I_{12}$ and $I_{14},I_{13}$.
This multiplicity can be nicely explained by the more refined Mondrian characterization, as we have
discussed with figure \ref{fig-16}.

\subsection{Anomalies and non-separable permutations}

At 5-loop, we have 15 anomalies out of $5!/2=60$ inequivalent combinations of ordered subspaces of $y$,
as given by the following list:
\be
Y(13524),~Y(14253),~Y(24135),~Y(31425), \labell{eq-8}
\ee
as well as
\be
\bal
&Y(23514),~Y(24153),~Y(24513),~Y(25134),~Y(25143),~Y(25314),\\
&Y(25413),~Y(31524),~Y(32514),~Y(35124),~Y(35214). \labell{eq-6}
\eal
\ee
They are separated into two parts since the first inherits anomalies from the 4-loop case, while the second
is the pure 5-loop contribution. Explicitly, this means we can rewrite
the 4 permutations of the first part ``separately'':
\be
Y(1,3524),~Y(1,4253),~Y(2413,5),~Y(3142,5),
\ee
which exhaust all four combinations of making a permutation separable at 5-loop but not at 4-loop. Note that
$(3524)$ is just a trivially shifted version of $(2413)$, as the rest two use their reverses.
For permutations of the second part, they are not separable at 5-loop already since in each of them, neither
1 or 5 stands at the boundary position.

To see their distinction more clearly, let's present all 15 expressions of 5-loop anomalies as
differences analogous to \eqref{eq-5}. First, the 4 separable pieces are given by
\be
A_{Y(13524)}=D_{12}D_{13}D_{14}D_{15}\times X_{23}X_{45}Y_{35}Y_{24}(X_{25}Y_{34}-X_{34}Y_{52}),
\ee
\be
A_{Y(14253)}=D_{12}D_{13}D_{14}D_{15}\times X_{23}X_{45}Y_{42}Y_{53}(X_{25}Y_{43}-X_{34}Y_{25}),
\ee
{}
\be
A_{Y(24135)}=D_{15}D_{25}D_{35}D_{45}\times X_{12}X_{34}Y_{24}Y_{13}(X_{14}Y_{23}-X_{23}Y_{41}),
\ee
\be
A_{Y(31425)}=D_{15}D_{25}D_{35}D_{45}\times X_{12}X_{34}Y_{31}Y_{42}(X_{14}Y_{32}-X_{23}Y_{14}),
\ee
in each of which the only effective part originates from the 4-loop anomaly \eqref{eq-5}, while
the rest factor is simply a trivial product of $D_{1j}$'s or $D_{i5}$'s.

Then, the 13 pure 5-loop contributions are given by
\be
\bal
A_{Y(23514)}=\,&\,X_{45}Y_{35}Y_{14}(X_{13}D_{24}D_{25}(D_{12}D_{23}-X_{12}X_{23})(X_{15}Y_{34}-X_{34}Y_{51})\\
&+X_{12}X_{23}Y_{34}D_{13}D_{25}(X_{15}Y_{24}-X_{24}Y_{51})+X_{12}X_{23}Y_{51}D_{13}D_{24}(X_{25}Y_{34}-X_{34}Y_{25})),
\eal
\ee
\be
\bal
A_{Y(25134)}=\,&\,X_{12}Y_{13}Y_{25}(X_{35}D_{24}D_{14}(D_{45}D_{34}-X_{45}X_{34})(X_{15}Y_{23}-X_{23}Y_{51})\\
&+X_{45}X_{34}Y_{23}D_{35}D_{14}(X_{15}Y_{24}-X_{24}Y_{51})+X_{45}X_{34}Y_{51}D_{35}D_{24}(X_{14}Y_{23}-X_{23}Y_{14})),
\eal
\ee
{}
\be
\bal
A_{Y(24153)}=\,&\,D_{25}(X_{12}X_{34}Y_{24}Y_{13}D_{15}(D_{45}D_{35}-Y_{45}Y_{53})(X_{14}Y_{23}-X_{23}Y_{41})\\
&+X_{13}X_{45}Y_{41}Y_{53}D_{23}(D_{12}D_{24}-X_{12}X_{24})(X_{15}Y_{43}-X_{34}Y_{15})),~~~~~~~~~~~~~~~~~~~~~~~~~~~
\eal
\ee
\be
\bal
A_{Y(31524)}=\,&\,D_{14}(X_{45}X_{23}Y_{24}Y_{35}D_{15}(D_{12}D_{13}-Y_{12}Y_{31})(X_{25}Y_{34}-X_{34}Y_{52})\\
&+X_{35}X_{12}Y_{52}Y_{31}D_{34}(D_{45}D_{24}-X_{45}X_{24})(X_{15}Y_{32}-X_{23}Y_{15})),~~~~~~~~~~~~~~~~~~~~~~~~~~~
\eal
\ee
{}
\be
\bal
A_{Y(24513)}=\,&\,X_{12}Y_{13}(X_{34}Y_{24}D_{15}D_{25}(D_{35}D_{45}-Y_{53}Y_{45})(X_{14}Y_{23}-X_{23}Y_{41})\\
&+X_{24}X_{35}Y_{25}Y_{45}D_{14}D_{34}(X_{15}Y_{23}-X_{23}Y_{51})+X_{14}X_{35}Y_{24}Y_{45}D_{23}D_{25}
(X_{15}Y_{43}-X_{34}Y_{51})),
\eal
\ee
\be
\bal
A_{Y(35124)}=\,&\,X_{45}Y_{35}(X_{23}Y_{24}D_{15}D_{14}(D_{13}D_{12}-Y_{31}Y_{12})(X_{25}Y_{34}-X_{34}Y_{52})\\
&+X_{24}X_{13}Y_{14}Y_{12}D_{25}D_{23}(X_{15}Y_{34}-X_{34}Y_{51})
+X_{25}X_{13}Y_{24}Y_{12}D_{34}D_{14}(X_{15}Y_{32}-X_{23}Y_{51})),
\eal
\ee
\be
\bal
A_{Y(25143)}=\,&\,X_{12}Y_{25}(X_{45}D_{35}(X_{34}Y_{13}Y_{23}D_{14}+Y_{43}D_{13}D_{23}Y_{14})(X_{15}Y_{24}-X_{24}Y_{51})\\
&+X_{34}X_{35}Y_{54}Y_{13}D_{14}D_{24}(X_{15}Y_{23}-X_{23}Y_{51})
+X_{34}X_{45}Y_{51}Y_{13}D_{24}D_{35}(X_{14}Y_{23}-X_{23}Y_{14})),
\eal
\ee
\be
\bal
A_{Y(32514)}=\,&\,X_{45}Y_{14}(X_{12}D_{13}(X_{23}Y_{35}Y_{34}D_{25}+Y_{32}D_{35}D_{34}Y_{25})(X_{15}Y_{24}-X_{24}Y_{51})\\
&+X_{23}X_{13}Y_{21}Y_{35}D_{25}D_{24}(X_{15}Y_{34}-X_{34}Y_{51})
+X_{23}X_{12}Y_{51}Y_{35}D_{24}D_{13}(X_{25}Y_{34}-X_{34}Y_{25})),
\eal
\ee
\be
\bal
A_{Y(25413)}=\,&\,X_{12}X_{34}Y_{13}(Y_{24}D_{15}D_{35}(D_{25}D_{45}-Y_{25}Y_{54})(X_{14}Y_{23}-X_{23}Y_{41})\\
&+X_{14}Y_{25}Y_{54}D_{24}D_{35}(X_{15}Y_{23}-X_{23}Y_{51})+X_{23}Y_{25}Y_{54}D_{15}D_{24}(X_{14}Y_{53}-X_{35}Y_{41})),~~
\eal
\ee
\be
\bal
A_{Y(35214)}=\,&\,X_{45}X_{23}Y_{35}(Y_{24}D_{15}D_{13}(D_{14}D_{12}-Y_{14}Y_{21})(X_{25}Y_{34}-X_{34}Y_{52})\\
&+X_{25}Y_{14}Y_{21}D_{24}D_{13}(X_{15}Y_{34}-X_{34}Y_{51})
+X_{34}Y_{14}Y_{21}D_{15}D_{24}(X_{25}Y_{31}-X_{13}Y_{52})),~~
\eal
\ee
{}
\be
\bal
A_{Y(25314)}=-X_{12}X_{23}X_{34}X_{45}Y_{25}Y_{53}Y_{31}Y_{14}D_{15}D_{24},
~~~~~~~~~~~~~~~~~~~~~~~~~~~~~~~~~~~~~~~~~~~~~~~~~~~~~~~~ \labell{eq-7}
\eal
\ee
where we have reorganized the order of 13 pieces in \eqref{eq-6} as they can be further interconnected:
$Y(25134)$ can be obtained from $Y(23514)$, by first reversing $(23514)$ to get $(41532)$ then interchanging
$1\!\leftrightarrow\!5$, $\!2\leftrightarrow\!4$, which explains the structural similarity between
$A_{Y(23514)}$ and $A_{Y(25134)}$ is no coincidence. It is the same for $Y(24153)$ and $Y(31524)$,
$Y(24513)$ and $Y(35124)$, $Y(25143)$ and $Y(32514)$, as well as $Y(25413)$ and $Y(35214)$.
The only special case is $Y(25314)$, which is invariant under the operation above.

\begin{figure}
\begin{center}
\includegraphics[width=0.6\textwidth]{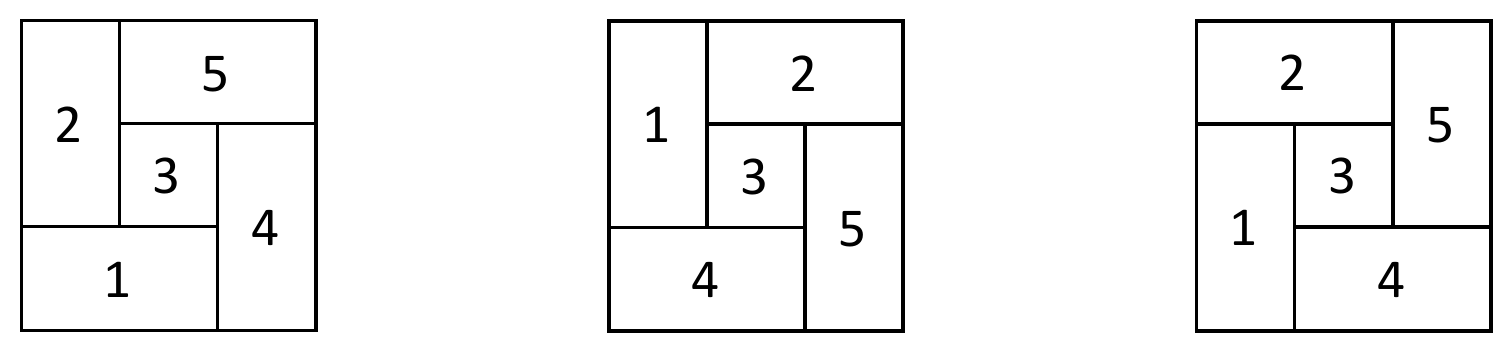}
\caption{Two clockwise and one counterclockwise pinwheel diagrams in subspace $Y(25314)$.} \label{fig-30}
\end{center}
\end{figure}

A more surprising property of $Y(25314)$ is, it can admit two clockwise pinwheel diagrams! It is for the
first time that a Mondrian configuration admits more than one way of filling numbers up to 5-loop, as shown in
figure \ref{fig-30}. This peculiarity of $Y(25314)$ explains why in anomaly \eqref{eq-7}
there are only terms with minus signs,
while in all other anomalies the numbers of terms with plus and minus signs are always equal. The latter fact reduces to
the complete cancelation if the particular permutation becomes separable, which is the completeness relation
in other words.

In figure \ref{fig-21}, as a specific example, we have demonstrated the uniqueness of number filling for ladders.
Then for a cross or a brick-wall, it is always possible to chop it into ladders, and repeat the filling process
for ladders analogous to that in figure \ref{fig-11}. However, for a pinwheel or a spiral diagram in general,
it is much more tricky to fill the numbers in a way consistent with two ordered subspaces simultaneously.

\begin{figure}
\begin{center}
\includegraphics[width=0.94\textwidth]{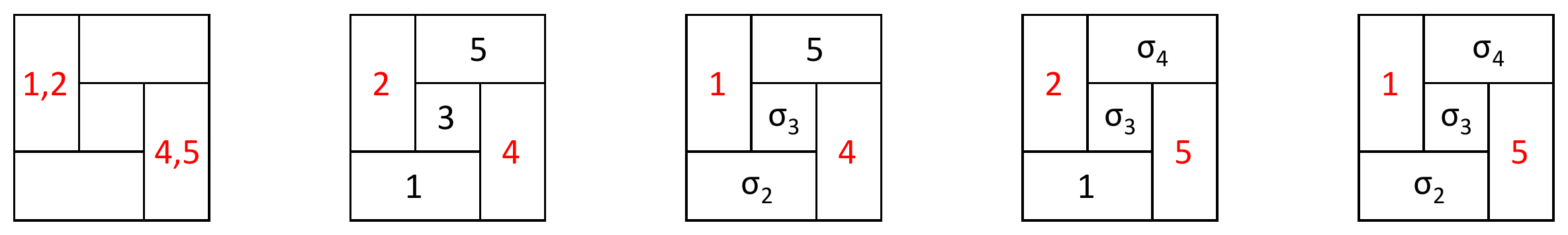}
\caption{Number filling of a clockwise pinwheel.} \label{fig-31}
\end{center}
\end{figure}

\begin{figure}
\begin{center}
\includegraphics[width=0.6\textwidth]{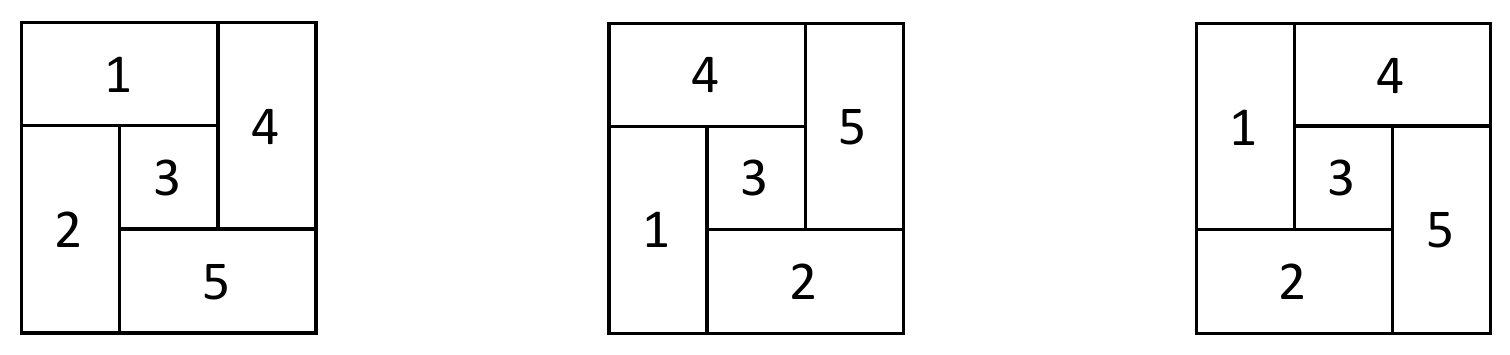}
\caption{Two counterclockwise and one clockwise pinwheel diagrams in subspace $Y(41352)$.} \label{fig-32}
\end{center}
\end{figure}

We present a simplest example of this process which is already complicated enough, for the clockwise pinwheel
in the left of figure \ref{fig-31}. For the upper left box, since there are three boxes on its left, its number
must be either 1 or 2, otherwise this is not consistent with $X(12345)$. The same logic applies for the lower right box,
of which the number must be either 4 or 5. Then they have four combinations, as given by the rest four diagrams
in figure \ref{fig-31}. If we choose 2 and 4 for these two positions, of the rest three blank boxes
the number in the top one must be 5 and the bottom one must be 1, which uniquely fixes its configuration of numbers.
If we choose 1 and 4, we can only fix the position of 5, then in the rest two boxes we can put $2,3$ arbitrarily,
and it is analogous for the choice of 2 and 5. Finally if we choose 1 and 5, none of the rest three can be fixed.
For these four cases, the numbers of admitted configurations are $1,2,2,6$ respectively.
Next, each configuration corresponds to 11 ordered subspaces of $y$, for example the first choice above
(the second diagram in figure \ref{fig-31}) admits
\be
\bal
&Y(25431),~Y(25341),~Y(25314),\\
&Y(54231),~Y(52431),~Y(52341),~Y(52314),\\
&Y(54321),~Y(53421),~Y(53241),~Y(53214),
\eal
\ee
as one can straightforwardly check. Therefore, there are in total $(1\!+\!2\!+\!2\!+\!6)\!\times\!11\!=\!121$
admitted subspaces of $y$, but there are only $5!\!=\!120$ combinations of them! This means at least one subspace
appears twice. A thorough check confirms it, as each of the rest 119 subspaces appears only once for one configuration,
but $Y(25314)$ appears twice, as we have known in figure \ref{fig-30}.
The clockwise pinwheel seems to be more special than the counterclockwise one,
but don't forget the reverse of $Y(25314)$! In the latter, namely $Y(41352)$, we have two counterclockwise pinwheels
instead, as shown in figure \ref{fig-32}, which is nothing but the reversed version of figure \ref{fig-30}
in the $y$ direction.

\begin{figure}
\begin{center}
\includegraphics[width=0.8\textwidth]{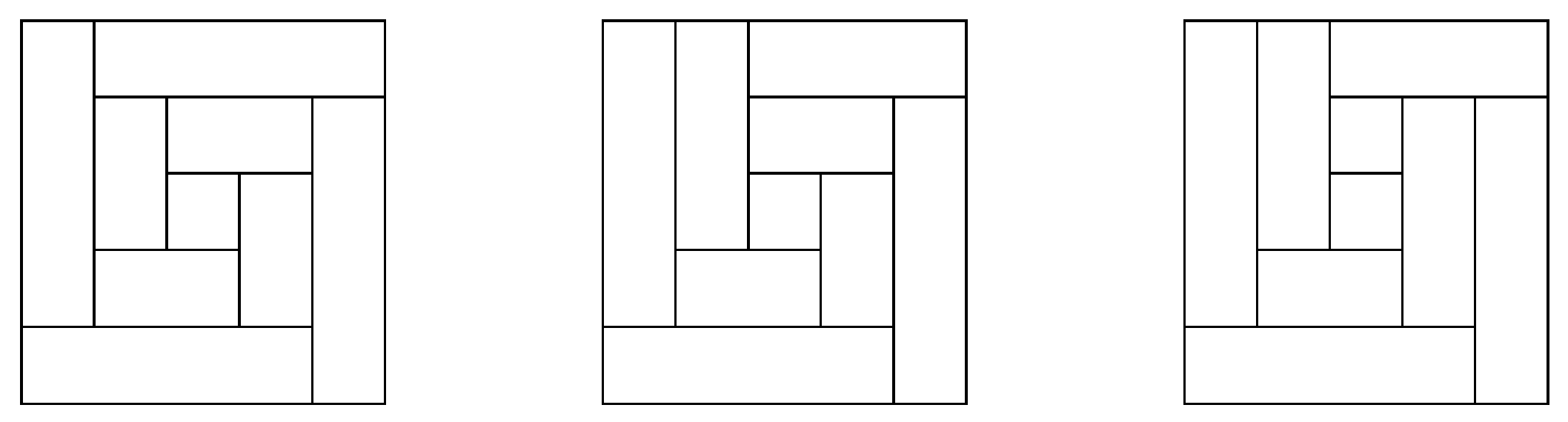}
\caption{Various topologies involving the S-pattern at 9-loop.} \label{fig-33}
\end{center}
\end{figure}

It is easy to imagine how complicated the number filling of a more general spiral diagram could be
as this process must be simultaneously consistent with the ordered subspaces of $x$ and $y$, unlike the available
step-by-step decomposition for ladders, crosses and brick-walls, as demonstrated by the example
in figure \ref{fig-21}. As a digression of curiosity
we present three more examples of spiral topologies in figure \ref{fig-33},
including a ``nested spiral'', a ``double spiral'' and a ``ladder in spiral'', among all
involving the S-pattern at 9-loop, in which the number filling is obviously more tricky.
For each of these diagrams, in general there would be more subspaces that admit more than one
configuration of numbers for one orientation.

Now we have comprehensively understood the 15 anomalies and non-separable permutations at 5-loop.
From now on let's define the 4 permutations non-separable at 4-loop only, as \textit{4-loop-non-separable} ones.
In appendix \ref{app1}, we further present the list of 163 non-separable permutations at 6-loop,
which are categorized into 4-, 5- and 6-loop-non-separable permutations according to the distinction above.

Finally it is worth emphasizing that, we have widened the definition of a non-separable permutation to make it
equivalent to the existence of an anomaly, as its complement is the \textit{totally} separable permutation.
As a contrast, we first present one example for each of 4-, 5- and 6-loop-non-separable permutations:
\be
(124635)\to(12)(4635),~~(134625)\to(1)(34625),~~(241563),
\ee
where the first two are partly separable, until this process fails at some point.
But for a totally separable permutation, $(153426$) for example, we have
\be
(153426)\to(1)(5342)(6)\to(1)(5)(34)(2)(6)\to(1)(5)(3)(4)(2)(6),
\ee
which can be finally separated as a row of single numbers. And only in such a case there is no anomaly, so that
the completeness relation holds. In general, for an $A$-loop-non-separable permutation at $L$-loop, with $A\!\leq\!L$,
$A$ is the cardinality of the maximal sub-permutation that is no longer partly separable.

\newpage
\section{All-loop Recursive Proof of the Completeness Relation}
\label{sec4}

Based on the 4-loop and 5-loop investigations in previous two sections, we have noticed sufficient evidence
for an all-loop recursive proof of the completeness relation. For an arbitrary separable permutation at $L$-loop,
its effectively contributing Mondrian diagrams used in this proof
can be always captured by those of the simplest subspace $Y(12\ldots L)$, while the rest allowed diagrams will
sum to zero due to the cancelation mechanism between C- and B-pattern. Hence this problem is separated into two parts,
which are nothing but the generalization of completeness relations for $Y(1234)$ and $Y(1324)$ at 4-loop, or those for
$Y(12345)$ and $Y(14325)$ at 5-loop, as we have previously done.

Typically, an all-loop recursive proof is achieved by showing that the $(L\!+\!1)$-loop case holds provided
the $L$-loop case holds. But due to the specific definition of separable permutations, we must generalize it
to show that the $(L_M\!+\!L_N)$-loop case holds provided the $L_M$- and $L_N$-loop ones hold, corresponding to
a permutation that can be separated into two sub-permutations which are also separable.
Now we consider a separable permutation or an ordered subspace $Y(\sg(M),\sg(N))$,
of which any number in set $M$ is always smaller than that in $N$, and both $\sg(M)$ and $\sg(N)$ are separable.
For example, $Y(4321\,765)$ is one of this category,
while $Y(2413\,765)$ is not since its sub-permutation $(2413)$ is not separable.

\begin{figure}
\begin{center}
\includegraphics[width=1.0\textwidth]{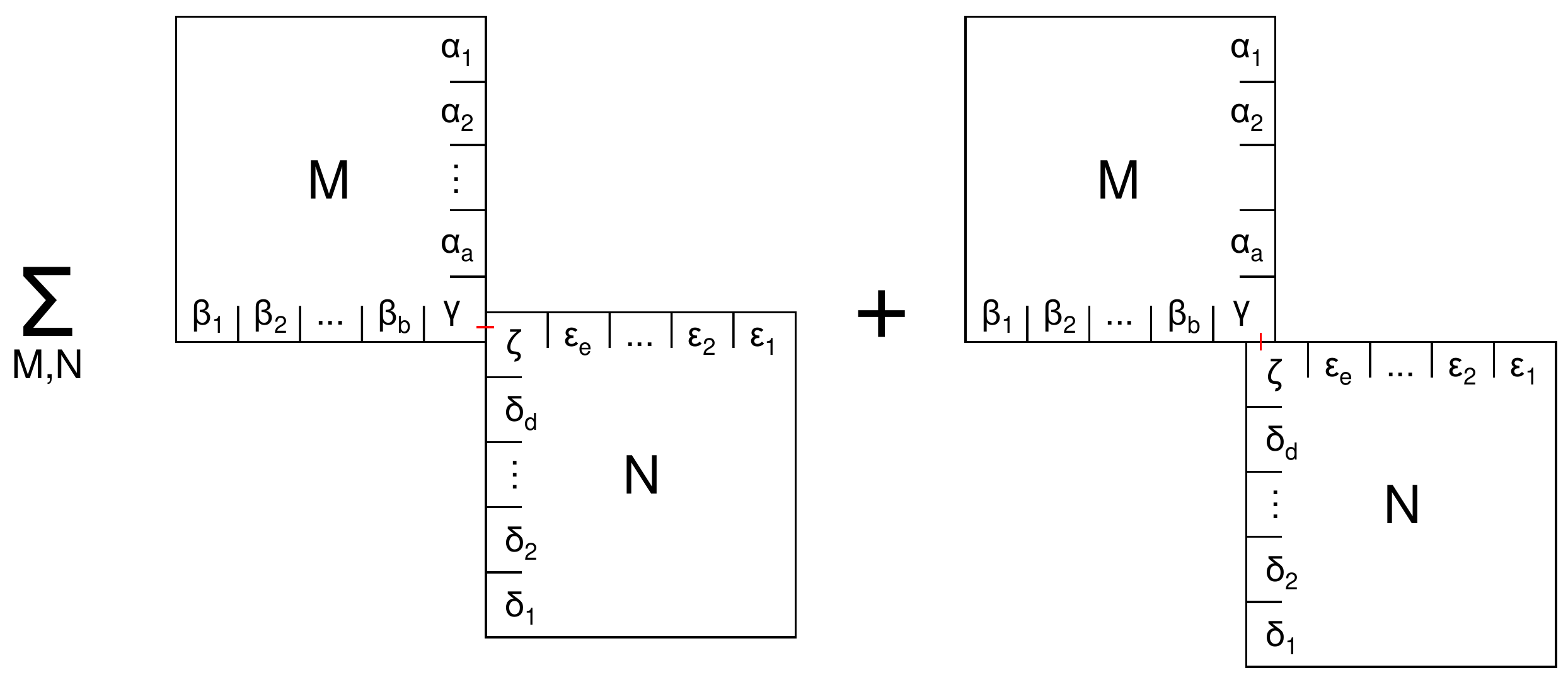}
\caption{Constructing the completeness relation for $Y(\sg(M),\sg(N))$ from those for $Y(\sg(M))$ and $Y(\sg(N))$.}
\label{fig-34}
\end{center}
\end{figure}

In figure \ref{fig-34}, we use the completeness relations for $Y(\sg(M))$ and $Y(\sg(N))$ to construct
the larger one for $Y(\sg_L)\!=\!Y(\sg(M),\sg(N))$ with $L\!=\!L_M\!+\!L_N$. Explicitly, we would like to prove
\be
\prod_{L}D=\sum_{M+N}\textrm{(Mondrian factor)} \labell{eq-9}
\ee
by judiciously combining
\be
\prod_{L_M}D=\sum_{M}\textrm{(Mondrian factor)},~\prod_{L_N}D=\sum_{N}\textrm{(Mondrian factor)}.
\ee
It is crucial to notice that, the box at the SE corner of each Mondrian diagram
which belongs to $M$ must have a contact with the box at the NW corner of each one which belongs to $N$. This fact is
highlighted in figure \ref{fig-34}, where boxes $\gm$ and $\zt$ have a contact that can be either horizontal or vertical,
otherwise when they have no contact, we get a NE-SW positioning of $\gm$ and $\zt$ which violates the fact that
$Y(\sg(M),\sg(N))$ is separable, as one can verify from the simplest examples in figures \ref{fig-10} and \ref{fig-11}.
This is a subtle but pivotal observation for the recursive proof below.

\subsection{Ordered subspace $X(12\ldots L)Y(12\ldots L)$}

To set aside the diagrams containing C- and B-pattern which may lead to cancelation for the moment, we can
choose $\sg(M)$ and $\sg(N)$ to be identities, namely $(12\ldots L_M)$ and $(L_M\!+\!1\,L_M\!+\!2\ldots L_M\!+\!L_N)$,
so that $Y(\sg(M),\sg(N))$ is trivially $Y(12\ldots L_M\!+\!L_N)$. Under this simplification, all diagrams in the
double sum in figure \ref{fig-34} have plus signs
and for diagrams containing B-patterns, the pair of non-contacting boxes can only have a NW-SE positioning,
while the exact cancelation between C- and B-pattern (with at least one NE-SW positioning
of non-contacting boxes for the latter) will be discussed later.

\begin{figure}
\begin{center}
\includegraphics[width=0.25\textwidth]{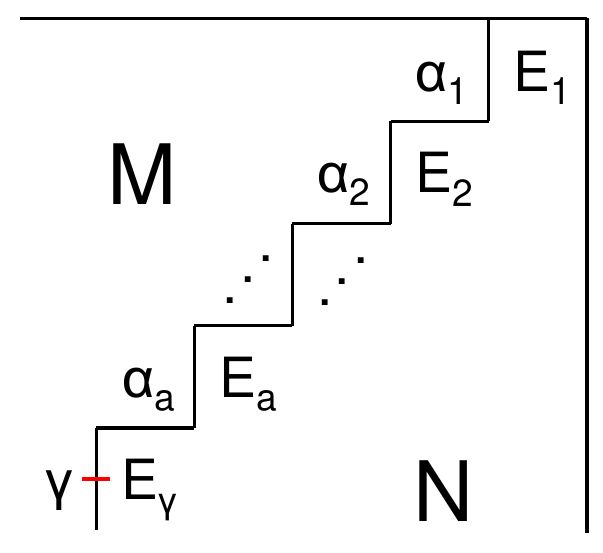}
\caption{Gluing the east boundary of $M$ and the north of $N$.} \label{fig-35}
\end{center}
\end{figure}

\begin{figure}
\begin{center}
\includegraphics[width=1.05\textwidth]{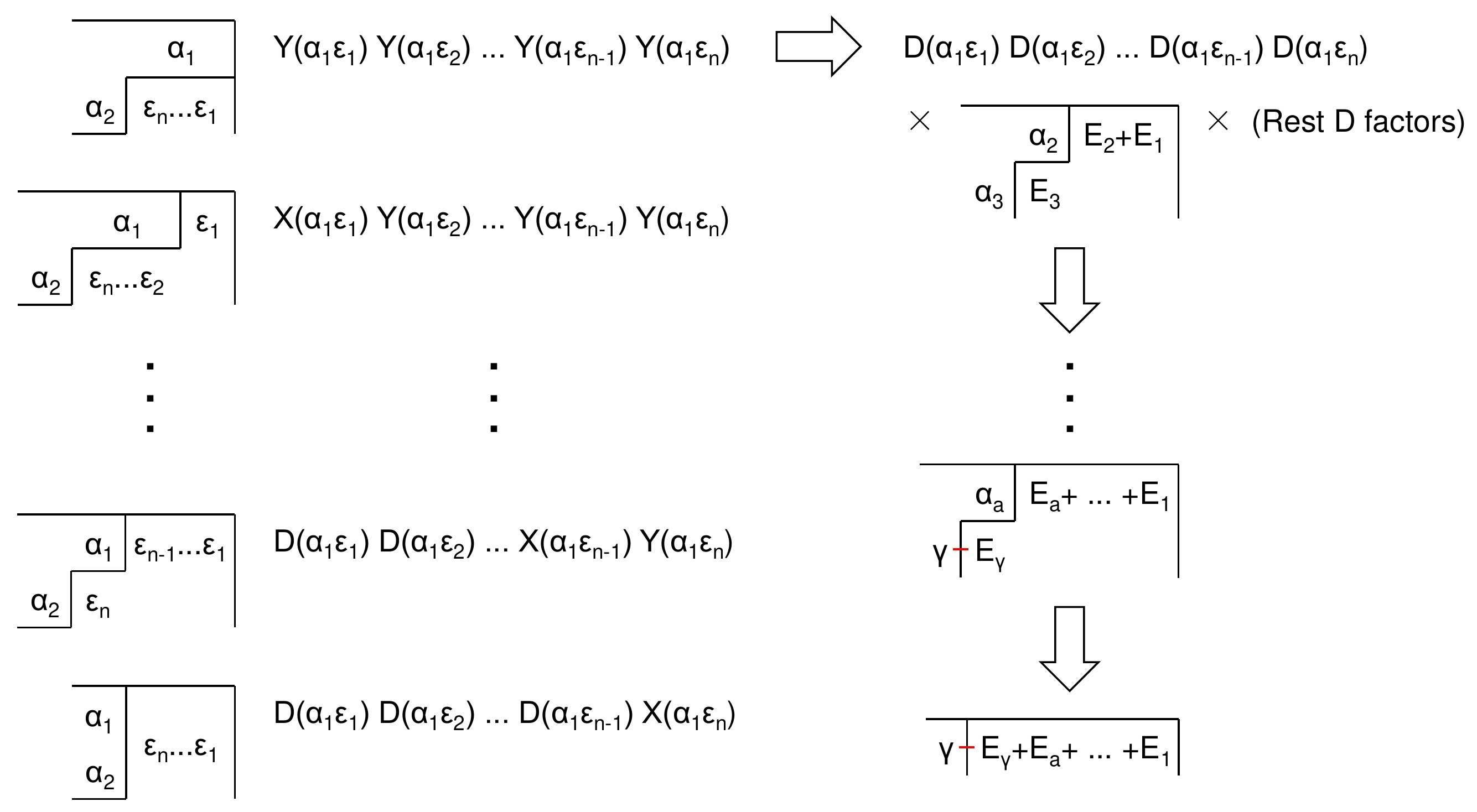}
\caption{Summing all possible partitions for the gluing of boundaries of $M,N$.} \label{fig-36}
\end{center}
\end{figure}

Now we will glue two diagrams which belong to $M$ and $N$ into a larger one for $(M\!+\!N)$, in the double sum
over $M,N$ in figure \ref{fig-34}. We may focus on the left diagram in which $\gm$ and $\zt$ have a horizontal contact,
more specifically, we focus on the gluing of the east boundary of $M$ and the north of $N$, as the rest cases
are analogous. The gluing is demonstrated in figure \ref{fig-35} where their contacting line is
a zig-zag path along the NE-SW direction, and we have defined sets $E_1,\ldots,E_a,E_\gm$ of which the union is
\be
E_1+\ldots+E_a+E_\gm=\{\ep_1,\ldots,\ep_e,\zt\},
\ee
where $\ep_1,\ldots,\ep_e,\zt$ are the labels of boxes along the north boundary of $N$ in figure \ref{fig-34}.
Note that each $E_i$ can also be empty for covering all possible partitions of $\{\ep_1,\ldots,\ep_e,\zt\}$.
Let's further zoom in to consider partitions $E_1,E_2$ adjacent to $\ap_1,\ap_2$ respectively in the NE corner of
figure \ref{fig-35}, then assume the union of $E_1,E_2$ is fixed, namely $E_1\!+\!E_2\!=\!\{\ep_1,\ldots,\ep_n\}$,
now we can begin the explicit enumeration shown in figure \ref{fig-36}. In the left column of figure \ref{fig-36},
we start with $E_1\!=\!\varnothing$ and end with $E_1\!=\!\{\ep_1,\ldots,\ep_n\}$, but in the latter case $\ap_2$ also
has a horizontal contact with $\ep_n$, which is inevitable, since $\ap_2$ sits below $\ap_1$ while there is no further
rightmost box other than $\ep_n$ as we have fixed $(E_1\!+\!E_2)$.

Then the sum of relevant Mondrian factors reads ($Y(ij)$ is used to replace $Y_{ij}$ and so forth)
\be
\bal
&\,Y(\ap_1\ep_1)\ldots Y(\ap_1\ep_n)+X(\ap_1\ep_1)Y(\ap_1\ep_2)\ldots Y(\ap_1\ep_n)\\
&+D(\ap_1\ep_1)X(\ap_1\ep_2)Y(\ap_1\ep_3)\ldots Y(\ap_1\ep_n)
+D(\ap_1\ep_1)D(\ap_1\ep_2)X(\ap_1\ep_3)Y(\ap_1\ep_4)\ldots Y(\ap_1\ep_n)\\
&+\ldots+D(\ap_1\ep_1)\ldots D(\ap_1\ep_{n-2})X(\ap_1\ep_{n-1})Y(\ap_1\ep_n)+D(\ap_1\ep_1)
\ldots D(\ap_1\ep_{n-1})X(\ap_1\ep_n)\\
=\,&\,D(\ap_1\ep_1)\ldots D(\ap_1\ep_n),
\eal
\ee
which is an overall factor on top of the configuration given in the top of the right column of figure \ref{fig-36},
and there are other trivial $D$ factors which are also universal. The key point here is to combine all factors with
$X$'s and $Y$'s \textit{along} the zig-zag contacting line between $M$ and $N$, to get an overall product of $D$'s,
while the factors with $X$'s and $Y$'s within $M$ or $N$ alone have been combined in the corresponding completeness
relation individually.

Completing the case of fixing $(E_1\!+\!E_2)$, we can proceed to
fix $(E_1\!+\!E_2\!+\!E_3)$, and so forth. The final configuration is given in the bottom of the right column of
figure \ref{fig-36}, and the same procedure can be done
for gluing $\{\bt_1,\ldots,\bt_b,\gm\}$ and $\{\de_1,\ldots,\de_d,\zt\}$,
as well as for those of the right diagram in figure \ref{fig-34} in which $\gm$ and $\zt$ have a vertical contact.
Finally, we are left with the only nontrivial sum $X(\gm\,\zt)\!+\!Y(\gm\,\zt)\!=\!D(\gm\,\zt)$
along the zig-zag contacting line, while all the rest factors have been turned into $D$'s.
Therefore, we have schematically achieved
\be
\sum_{M,N}\sum_{\ap\textrm{-}\ep,\,\bt\textrm{-}\de}
\textrm{(Mondrian factor)}_\textrm{H}+\textrm{(Mondrian factor)}_\textrm{V}=\prod_{L}D,
\ee
where $\ap$-$\ep$ and $\bt$-$\de$ denote all possible contacting lines between
$\{\ap_1,\ldots,\ap_a,\gm\}$ and $\{\ep_1,\ldots,\ep_e,\zt\}$, as well as $\{\bt_1,\ldots,\bt_b,\gm\}$ and $\{\de_1,\ldots,\de_d,\zt\}$. And H or V denotes the relation between $\gm$ and $\zt$ indicated in figure \ref{fig-34}.
On the other hand, the LHS above covers all possible Mondrian diagrams in subspace $Y(12\ldots L)$ as
\be
\sum_{M,N}\sum_{\ap\textrm{-}\ep,\,\bt\textrm{-}\de}
\textrm{(Mondrian factor)}_\textrm{H}+\textrm{(Mondrian factor)}_\textrm{V}=\sum_{M+N}\textrm{(Mondrian factor)},
\ee
which can been seen more transparently in a reverse way, that is, for a given $(L_M\!+\!L_N)$-loop diagram and
a fixed partition of $M$ and $N$, the zig-zag contacting line which cuts it into two parts is unique,
so are the two resulting sub-diagrams. In other words, if we sum over all possible combinations of sub-diagrams
and contacting lines which are exactly done in the sum above,
we obtain all possible Mondrian diagrams. This completes the recursive proof of
completeness relation \eqref{eq-9} for $Y(12\ldots L)$.

One remark is, there is no cross along the zig-zag contacting line as well as within $M$ or $N$.
We know that it is not allowed in $Y(12\ldots L_M\!+\!L_N)$,
but essentially it is the gluing that forbids it, as having a cross along that line will make properly
``shrinking'' $M$ or $N$ back to its original shape in figure \ref{fig-34} impossible.

\subsection{Exact cancelation between C- and B-pattern}

\begin{figure}
\begin{center}
\includegraphics[width=0.35\textwidth]{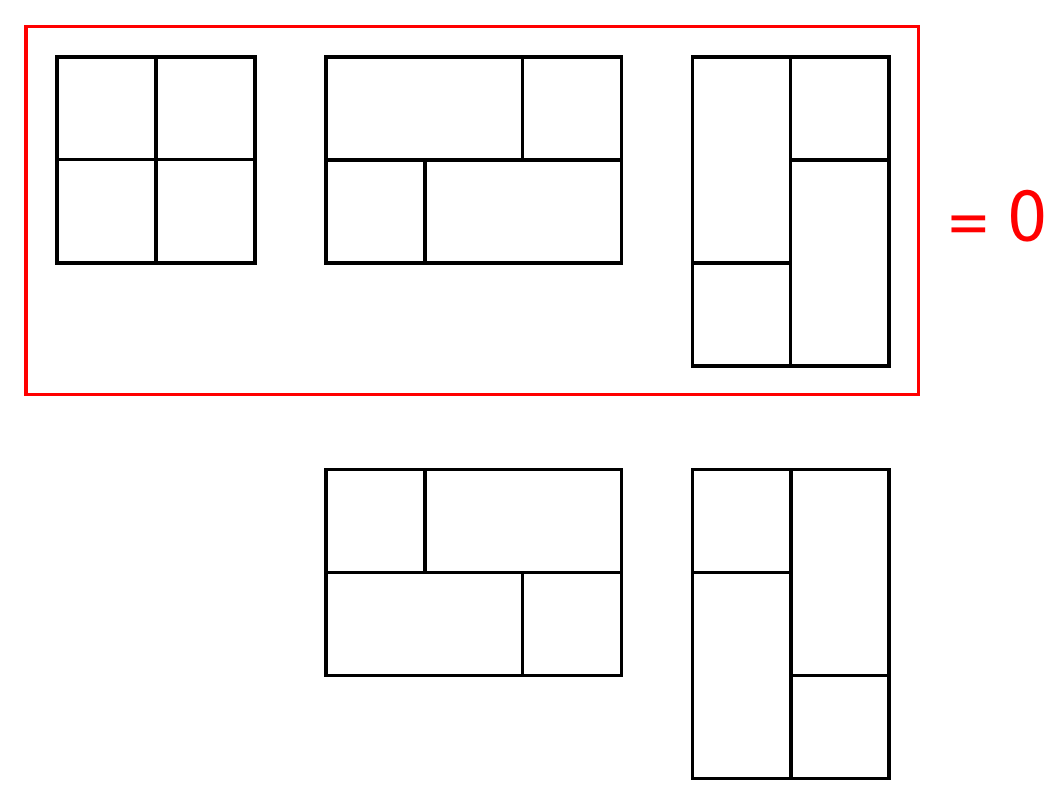}
\caption{Cancelation between a cross and a pair of NE-SW brick-walls.} \label{fig-37}
\end{center}
\end{figure}

Next we will demonstrate that, at $L$-loop, in subspace $Y(\sg_L)$ where $\sg_L$ is an arbitrary separable permutation,
the effectively contributing Mondrian diagrams are in fact identical to those in $Y(12\ldots L)$ except for the
configurations of numbers that depend on the ordering $\sg_L$. This is due to the fact that,
the cancelling diagrams containing C- and B-pattern always appear simultaneously so that as if they were absent.
As we have shown for $Y(1234)$ and $Y(1324)$ at 4-loop, its mechanism can be essentially captured by the simplest
cancelation between a cross and a pair of brick-walls given in figure \ref{fig-37}.
There, all explicit configurations of numbers have been suppressed since they don't really matter,
and we always choose the pair of brick-walls in each of which the pair of non-contacting boxes has a NE-SW positioning
(from now on, let's call them NE-SW brick-walls or B-patterns for short)
to cancel the cross while preserving the NW-SE ones.

\begin{figure}
\begin{center}
\includegraphics[width=0.24\textwidth]{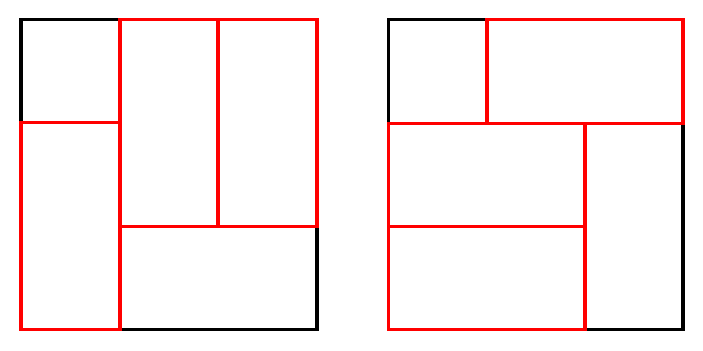}
\caption{A chain of boxes that align either horizontally or vertically.} \label{fig-38}
\end{center}
\end{figure}

Recall that for B-patterns, a pair of non-contacting boxes are those cannot be
connected by a chain of boxes that align either horizontally or vertically. In contrast, as examples,
non-contacting boxes (highlighted in red) given in figure \ref{fig-38} can be connected by such a chain of boxes,
hence they are not non-contacting in the sense of B-patterns, as they merely form deformed ladders essentially.

\begin{figure}
\begin{center}
\includegraphics[width=0.51\textwidth]{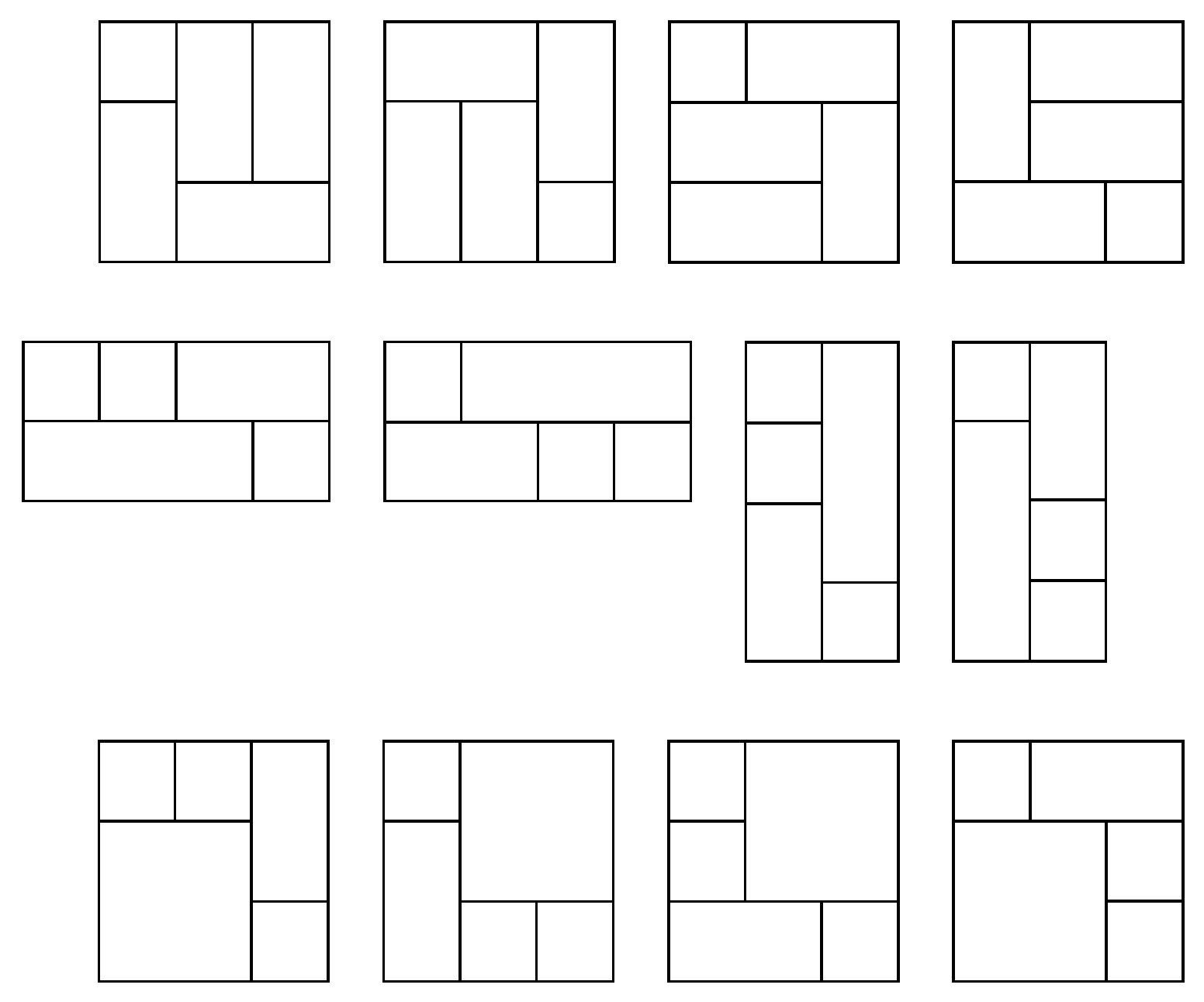}
\caption{Contributing Mondrian diagrams containing NW-SE B-patterns at 5-loop.} \label{fig-39}
\end{center}
\end{figure}

\begin{figure}
\begin{center}
\includegraphics[width=1.03\textwidth]{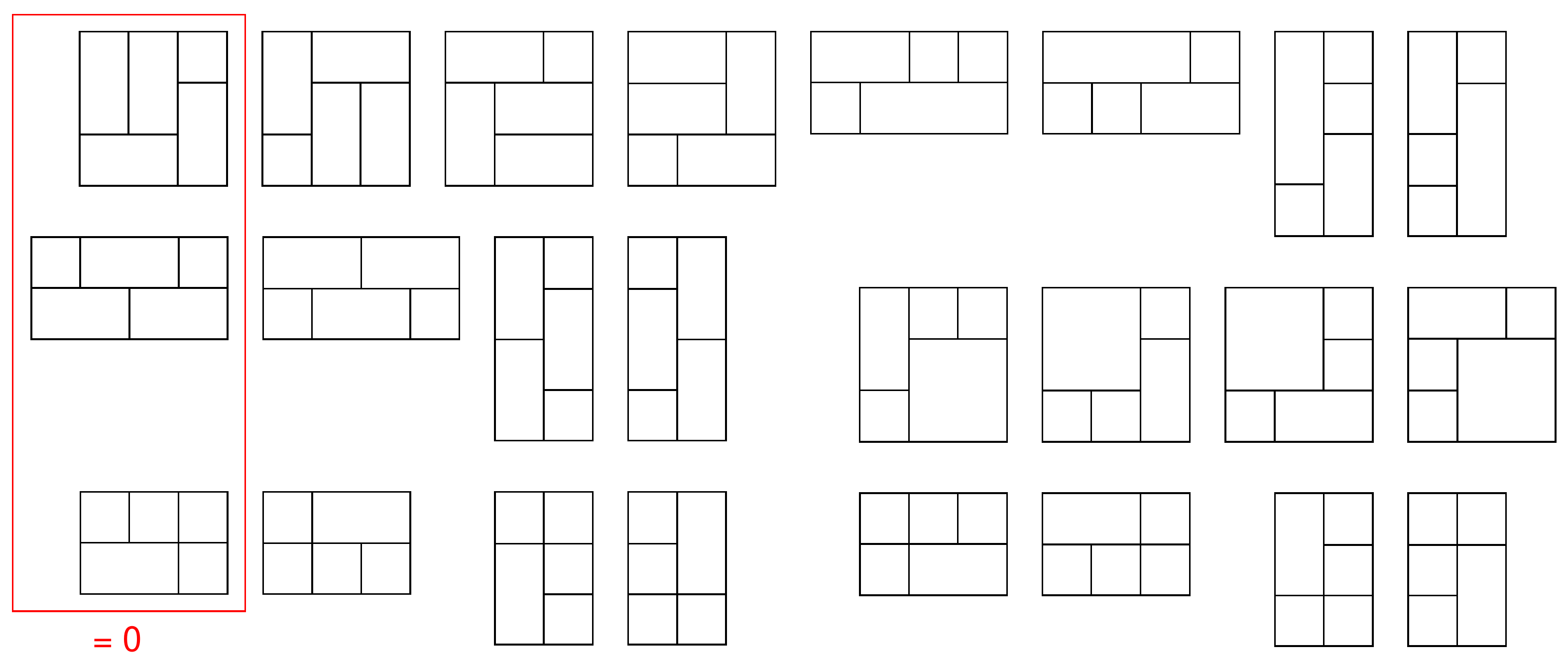}
\caption{Cancelling Mondrian diagrams containing C- and NE-SW B-patterns at 5-loop.} \label{fig-40}
\end{center}
\end{figure}

At 5-loop, we can find more nontrivial examples of the cancelation mechanism. In figures \ref{fig-39} and \ref{fig-40},
we reorganize all relevant Mondrian diagrams, and separate them into the contributing and the cancelling parts
containing C- and B-pattern regardless of in which subspace provided it is a separable permutation.
It is clear that all diagrams in figure \ref{fig-39} are NW-SE brick-walls, namely all those admitted
in $Y(12\ldots L)$. They represent half of all possible orientations of the three topologies they belong to,
while the rest half, namely the NE-SW brick-walls, are given in figure \ref{fig-40}.

The key point of this separation is, as dihedral symmetry indicates, in general there are eight orientations
which include: the topology itself, its images rotated by 180 degrees, reversed along the NW-SE and
NE-SW directions, as well as the left-right (or up-down) reflections of these four. Note that either the former
or latter four orientations preserve the NW-SE or NE-SW positioning for a pair of non-contacting boxes.

Following the cancelation in figure \ref{fig-37},
we choose the pairs of NE-SW brick-walls to cancel the crosses respectively, as indicated in figure \ref{fig-40}.
There, each of the eight columns of diagrams sum to zero,
so that all eight orientations of the cross topology can be exhausted.
Note the first four diagrams in the 2nd row represent a topology absent in figure \ref{fig-39},
as it has an additional symmetry of reflection and hence only four different orientations, in each of which both
NW-SE and NE-SW B-patterns appear.

It is a general fact that, all possible diagrams of such a topology always cancel with those containing C-patterns
since no matter how we orient them, the NE-SW B-patterns, which are deliberately chosen to cancel the C-patterns,
always exist. This manually designed mechanism plays a crucial role in generalizing the recursive proof of
completeness relation from that for $Y(12\ldots L)$ only, to that for $Y(\sg_L)$ provided
$\sg_L$ is a separable permutation. The reasoning is simple: for $Y(\sg_L)\!=\!Y(\sg(M),\sg(N))$,
due to the cancelation mechanism of $Y(\sg(M))$ and $Y(\sg(N))$, we are always left with
the contributing diagrams in $Y(12\ldots L_M)$ and $Y(L_M\!+\!1\,L_M\!+\!2\ldots L_M\!+\!L_N)$ respectively,
so we can schematically write
\be
\bal
\textrm{Diag}(\sg(M),\sg(N))&=\textrm{Diag}(\sg(M))\otimes\textrm{Diag}(\sg(N))\\
&=(\textrm{Diag}(12\ldots L_M)+0_M)\otimes(\textrm{Diag}(L_M\!+\!1\,L_M\!+\!2\ldots L_M\!+\!L_N)+0_N)\\
&=\textrm{Diag}(12\ldots L_M\!+\!L_N),
\eal
\ee
where `Diag' stands for all possible Mondrian diagrams, $\otimes$ the gluing of two sets of diagrams,
and `0' the cancelling diagrams respectively. Of course, we have suppressed all explicit configurations of numbers.
It is important to ensure that, all possible orientations of all Mondrian topologies
have been enumerated, so that we can reorganize them in a way
similar to that of figures \ref{fig-39} and \ref{fig-40}, to get a neat cancelation.
This completes the recursive proof of completeness relation \eqref{eq-9} for a general separable $Y(\sg_L)$.

We present a simplest nontrivial example: to recursively prove the completeness relation of $Y(1324)$
from that of $Y(1234)$. Straightforwardly, one can use
$\textrm{Diag}(1324)\!=\!\textrm{Diag}(132)\!\otimes\!\textrm{Diag}(4)$, or
$\textrm{Diag}(1324)\!=\!\textrm{Diag}(1)\!\otimes\!\textrm{Diag}(324)$, then return to $\textrm{Diag}(1234)$.
In contrast, $\textrm{Diag}(1324)\!=\!\textrm{Diag}(13)\!\otimes\!\textrm{Diag}(24)$ is illegal.

\begin{figure}
\begin{center}
\includegraphics[width=0.21\textwidth]{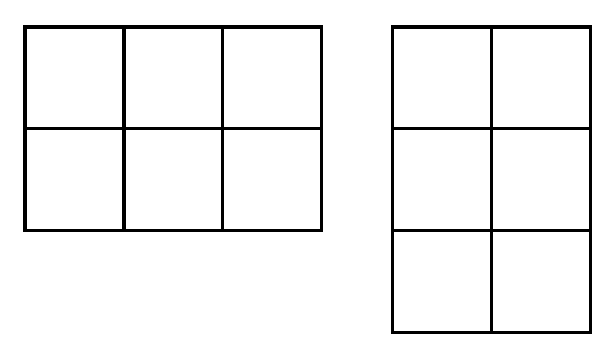}
\caption{All two possible orientations of the double-cross topology at 6-loop.} \label{fig-41}
\end{center}
\end{figure}

\begin{figure}
\begin{center}
\includegraphics[width=0.83\textwidth]{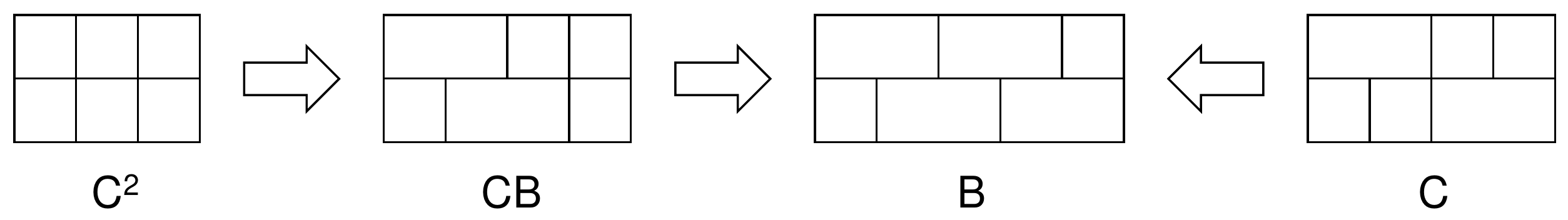}
\caption{A special brick-wall is shared by two families of topologies containing C$^2$ and C.} \label{fig-42}
\end{center}
\end{figure}

\begin{figure}
\begin{center}
\includegraphics[width=0.48\textwidth]{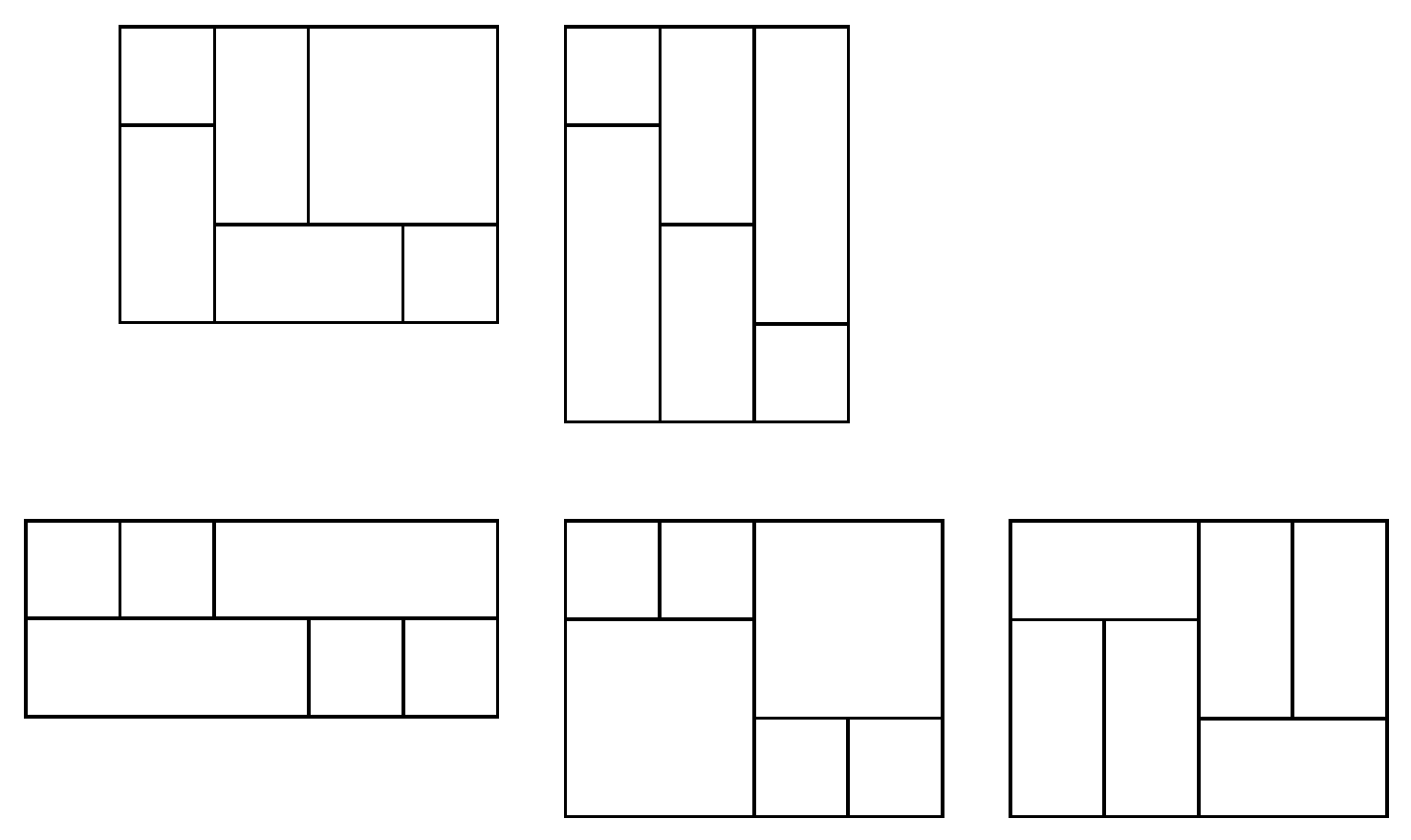}
\caption{Surviving topologies after the cancelation between two families containing C$^2$ and C.} \label{fig-43}
\end{center}
\end{figure}

Finally, let's see how the cancelation mechanism works for more than one cross or C-pattern, for this purpose
we must go beyond 5-loop. In figure \ref{fig-41}, we present a double-cross topology at 6-loop with all two
possible orientations, as a simplest example. Denoted by C$^2$, it has two derivative
CB topologies, and six derivative B topologies, note the power of C matters since it determines the sign of a topology.
Its entire family of $(1\!+\!2\!+\!6)$ topologies is given in appendix \ref{app2} where all Mondrian topologies at 6-loop
are listed in a classified way specifying the C-, B- and S-pattern. We may pick the left diagram in figure \ref{fig-41},
precisely denoted by $\textrm{C}\otimes_\textrm{H}\!\textrm{C}$ where $\otimes_\textrm{H}$ means the two crosses align
horizontally, for the following discussion. If we symbolically write the enumeration of cross and brick-wall diagrams
in figure \ref{fig-37} as
\be
-\,\textrm{C}+\textrm{H}'+\textrm{V}'+\textrm{H}+\textrm{V}=\textrm{H}+\textrm{V},
\ee
where H$'$ and V$'$ denote the horizontal and vertical NE-SW B-patterns, and H and V the NW-SE ones,
we then can enumerate all relevant diagrams with respect to $\otimes_\textrm{H}$ at 6-loop
(including $\textrm{C}\otimes_\textrm{H}\!\textrm{C}$ itself, its sign is $(-)^2\!=\!+$) as
\be
\bal
&(-\,\textrm{C}+\textrm{H}'+\textrm{V}'+\textrm{H}+\textrm{V})\otimes_\textrm{H}\!
(-\,\textrm{C}+\textrm{H}'+\textrm{V}'+\textrm{H}+\textrm{V})\\
=\,&(\textrm{H}+\textrm{V})\otimes_\textrm{H}\!(\textrm{H}+\textrm{V})\\
=\,&\textrm{H}\otimes_\textrm{H}\!\textrm{H}+\textrm{H}\otimes_\textrm{H}\!\textrm{V}
+\textrm{V}\otimes_\textrm{H}\!\textrm{H}+\textrm{V}\otimes_\textrm{H}\!\textrm{V},
\eal
\ee
which leave four surviving diagrams. Note that before we apply the cancelation
$-\textrm{C}\!+\!\textrm{H}'\!+\!\textrm{V}'\!=\!0$, there are $5^2\!=\!25$ diagrams
for $\otimes_\textrm{H}$, together with the other 25 ones for $\otimes_\textrm{V}$ we find that they
precisely reproduce the $2\!+\!(8\!+\!8)\!+\!(4\!+\!8\!+\!4\!+\!8\!+\!4\!+\!4)\!=\!50$ diagrams
in the entire family of this double-cross topology. However, not all four diagrams above survive another
cancelation indicated in figure \ref{fig-42}. The first diagram, namely $\textrm{H}\otimes_\textrm{H}\!\textrm{H}$,
is a brick-wall shared by two families of topologies containing C$^2$ and C respectively.
The entire family of $(1\!+\!4)$ topologies of the latter is also given in appendix \ref{app2}. As expected, the topology
of $\textrm{H}\otimes_\textrm{H}\!\textrm{H}$
contains both NW-SE and NE-SW B-patterns, and hence does not contribute at all.

In summary, the double-cross family at 6-loop effectively contributes
\be
\bal
\,&\ldots+\textrm{H}\otimes_\textrm{H}\!\textrm{V}+\textrm{V}\otimes_\textrm{H}\!\textrm{H}
+\textrm{V}\otimes_\textrm{H}\!\textrm{V}\\
\,&~~~~+\textrm{H}\otimes_\textrm{V}\!\textrm{V}+\textrm{V}\otimes_\textrm{V}\!\textrm{H}
+\textrm{H}\otimes_\textrm{V}\!\textrm{H},
\eal
\ee
which are exactly the $(4\!+\!2)$ orientations of the two topologies given in the 1st row of figure \ref{fig-43}
as $\textrm{V}\otimes_\textrm{V}\!\textrm{V}$ similarly does not contribute. And the 2nd row of figure \ref{fig-43}
gives the three surviving derivative topologies of the cross in the rightmost of figure \ref{fig-42}.
From this example we also learn that, while the C- and NE-SW B-pattern always appear simultaneously so that
they can neatly cancel, it is more nontrivial to keep track of those topologies shared by different families
for correctly identifying the final surviving diagrams.

\section{Imperfections to be Explored and Improved}

With the all-loop completeness relation, we can conveniently check whether all Mondrian topologies have been considered
for a particular ordered subspace, which means we must explicitly enumerate all possible orientations of them,
and also use the cancelation between C- and B-pattern when available. But so far, we do not know a more efficient way
to do this enumeration other than more or less using intuition. This is the most relevant imperfection to be explored
in Mondrian diagrammatics.

Another imperfection is related to the non-Mondrian diagrams as we have mentioned since the 4-loop case.
Can we generalize all these simple and nice mathematical machineries to include the non-Mondrian complexity,
do they have more nontrivial connections to dual conformal invariance, and finally what roles do they play in
the 4-particle amplituhedron or integrand? These are all very intriguing questions towards the ultimate goal:
to understand the amplituhedron from a combinatoric perspective, which turns out to be equivalent to the complete
triangulation of amplituhedron and hence frees it from the latter. This new and complete perspective must be able to
account for the existence of non-Mondrian diagrams, as well as various coefficients of their corresponding
dual conformally invariant integrals other than $\pm1$ (such as 0, $+2$ or even fractional ones) beyond 5-loop order.

Besides the big or ambitious aspects, the small technical imperfections also deserve attentions.
These include: the dual conformally invariant replacement $x_{ji}z_{ij}$ (or $y_{ji}w_{ij}$)$\,\to\!D_{ij}$
for each extra connecting line between two loops, and the integral convergence preserving replacement
$x_jz_i\!\to\!D_{ij}$ or $y_lw_k\!\to\!D_{kl}$, or both of them (with an unexplained minus sign in this case)
for the S-pattern. For the latter, we also discuss its application at 6-loop in appendix \ref{app2}, where there exist
more than one topology containing the S-pattern.

One final remark is, though we have checked the 7-loop completeness relation in a particular subspace which is chosen
to be $Y(1634527)$, after identifying and classifying all topologies that agree with \cite{Bourjaily:2011hi},
these ``data'' do not seem to have new features that are absent up to 6-loop, therefore we decide to relegate
the additional appendix listing all Mondrian topologies at 7-loop to a future edition (if necessary).

\section*{Acknowledgments}

This work is partly supported by Qiu-Shi Funding and Chinese NSF funding under contracts\\
No.11135006, No.11125523 and No.11575156.

\newpage
\appendix
\section{Non-separable Permutations at 6-loop}
\label{app1}

Before we list all 163 non-separable permutations at 6-loop, it is instructive to reorganize
those at 5-loop, namely \eqref{eq-8} and \eqref{eq-6}, then the 6-loop case will be its natural generalization.
Modulo the trivial reversion in the $y$ direction, we always choose the \textit{lexicographically} smaller permutation,
so that \eqref{eq-8} can be written more clearly as
\be
\bal
1\left\{\begin{array}{c}
3524 \\
4253
\end{array} \right.,~
\left.\begin{array}{c}
2413 \\
3142
\end{array} \right\}5\,,
\eal
\ee
note that 1 and 5 are placed in the front and end respectively. For \eqref{eq-6}, numbers in the front and end
can be chosen as
\be
\bal
2\,\underline{~~}\,\underline{~~}\,\underline{~~}\,3,~
&2\,\underline{~~}\,\underline{~~}\,\underline{~~}\,4,\\
&3\,\underline{~~}\,\underline{~~}\,\underline{~~}\,4,
\eal
\ee
then according to the definition of non-separable permutations, we can have
\be
\bal
\left\{\begin{array}{c}
24\,\sg(15)\,3 \\
25\,\sg(14)\,3
\end{array} \right.,
&~~~\,2\,\sg(3,51)\,4\,, \\
&\left\{\begin{array}{c}
3\,\sg(15)\,24 \\
3\,\sg(25)\,14
\end{array} \right.,
\eal
\ee
where $\sg(3,51)$ includes $351$, $531$ and $513$. These are the $(4\!+\!11)$ 4- and 5-loop-non-separable permutations
at 5-loop under this more systematic enumeration.

Then for the 6-loop case, we similarly start with the 4-loop-non-separable permutations, given by
\be
\bal
&\sg(12)\left\{\begin{array}{c}
4635 \\
5364
\end{array} \right.,~
1\left\{\begin{array}{c}
4635 \\
5364
\end{array} \right\}2,~
\left.\begin{array}{c}
2413 \\
3142
\end{array} \right\}\sg(56)\,,~
5\left\{\begin{array}{c}
2413 \\
3142
\end{array} \right\}6,\\
&~~~~16\left\{\begin{array}{c}
3524 \\
4253
\end{array} \right.,~
1\left\{\begin{array}{c}
3524 \\
4253
\end{array} \right\}6,~
\left.\begin{array}{c}
3524 \\
4253
\end{array} \right\}16\,,
\eal
\ee
and there are in total 18 such permutations. The 5-loop-non-separable parts are given by
\be
\bal
&1\left\{\begin{array}{c}
A(23456) \\
A(65432)
\end{array} \right.,~
\left.\begin{array}{c}
A(12345) \\
A(54321)
\end{array} \right\}6,
\eal
\ee
where $A(12345)$ includes the 13 5-loop-non-separable permutations in \eqref{eq-6}, while $A(54321)$ stands for its
reverse, and similar for $A(23456)$ and $A(65432)$. There are in total $11\!\times\!4\!=\!44$ such permutations.

For the 6-loop-non-separable parts, again, numbers in the front and end can be chosen as
\be
\bal
2\,\underline{~~}\,\underline{~~}\,\underline{~~}\,3,~
2\,\underline{~~}\,\underline{~~}\,\underline{~~}\,4,~
&2\,\underline{~~}\,\underline{~~}\,\underline{~~}\,5,\\
3\,\underline{~~}\,\underline{~~}\,\underline{~~}\,4,~
&3\,\underline{~~}\,\underline{~~}\,\underline{~~}\,5,\\
&4\,\underline{~~}\,\underline{~~}\,\underline{~~}\,5,
\eal
\ee
continuing the enumeration of non-separable permutations case by case, we can have
\be
\bal
2\left\{\begin{array}{c}
4\,\sg(156) \\
5\,\sg(146) \\
6\,\sg(145)
\end{array} \right\}3,~~
2\left\{\begin{array}{c}
35\,\sg(16) \\
36\,\sg(15) \\
5\,\sg(136) \\
6\,\sg(135)
\end{array} \right\}4,~~
&2\left\{\begin{array}{c}
3\,\sg(4,61) \\
4\,\sg(16)\,3~ \\
4\,\sg(36)\,1~ \\
6\,\sg(134)~\,
\end{array} \right\}5, \\
3\left\{\begin{array}{c}
\sg(15)\,\sg(26) \\
\sg(26)\,\sg(15) \\
\sg(16)\,\sg(25) \\
\sg(25)\,\sg(16) \\
\sg(56)\,\sg(12)
\end{array} \right\}4,~~
&3\left\{\begin{array}{c}
\sg(16)\,24 \\
\sg(26)\,14 \\
\sg(146)\,2 \\
\sg(246)\,1
\end{array} \right\}5, \\
&4\left\{\begin{array}{c}
\sg(126)\,3 \\
\sg(136)\,2 \\
\sg(236)\,1
\end{array} \right\}5,
\eal
\ee
and there are in total $18\!+\!16\!+\!13\!+\!20\!+\!16\!+\!18\!=\!101$ such permutations.

In summary, there are $18\!+\!44\!+\!101\!=\!163$ non-separable permutations
out of $6!/2\!=\!360$ combinations at 6-loop, and we have confirmed that only for these 163 subspaces of $y$,
the anomalies are nonzero.

\newpage
\section{All Mondrian Topologies at 6-loop}
\label{app2}

\begin{figure}
\begin{center}
\includegraphics[width=1.03\textwidth]{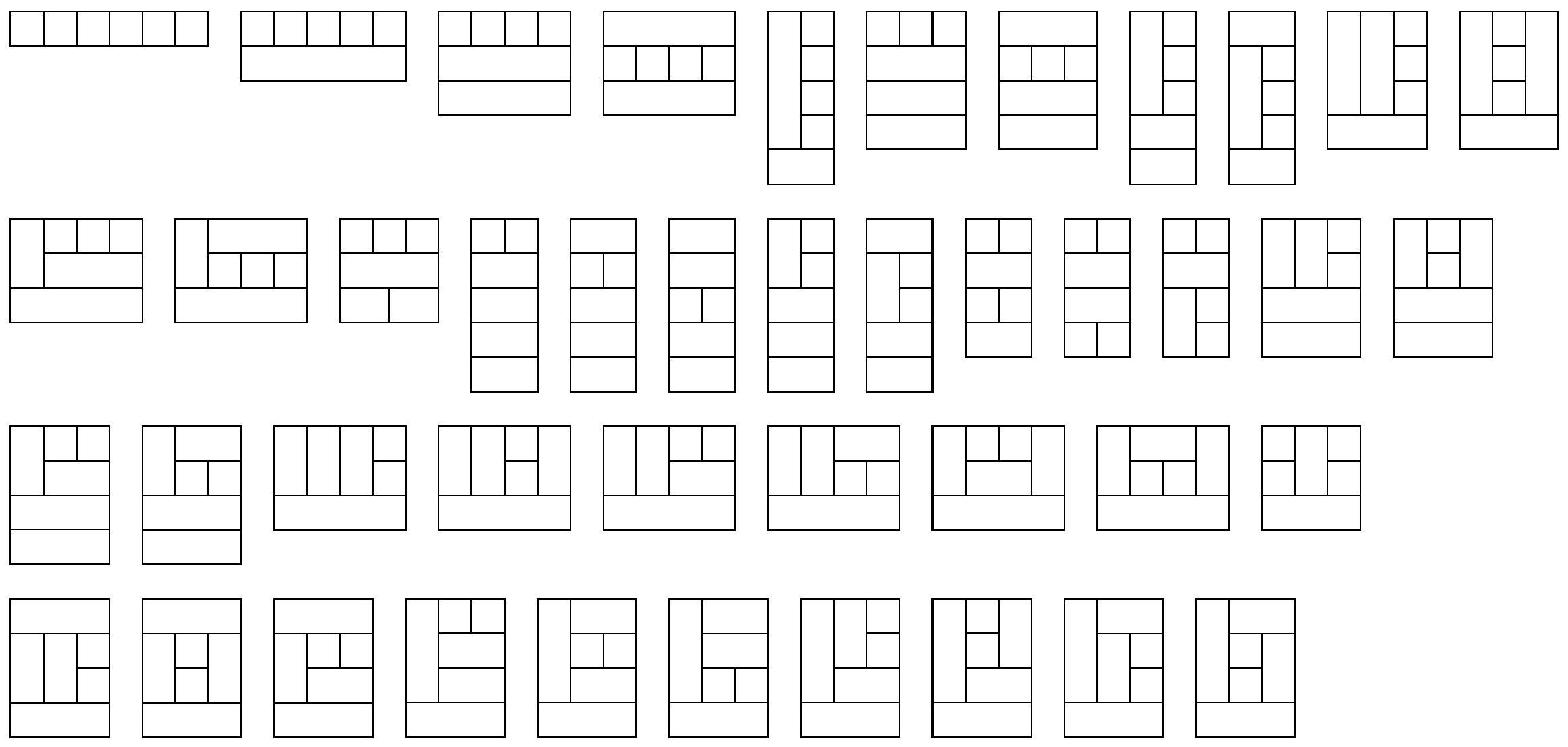}
\caption{Topologies of the L-pattern at 6-loop.} \label{fig-44}
\end{center}
\end{figure}

\begin{figure}
\begin{center}
\includegraphics[width=0.59\textwidth]{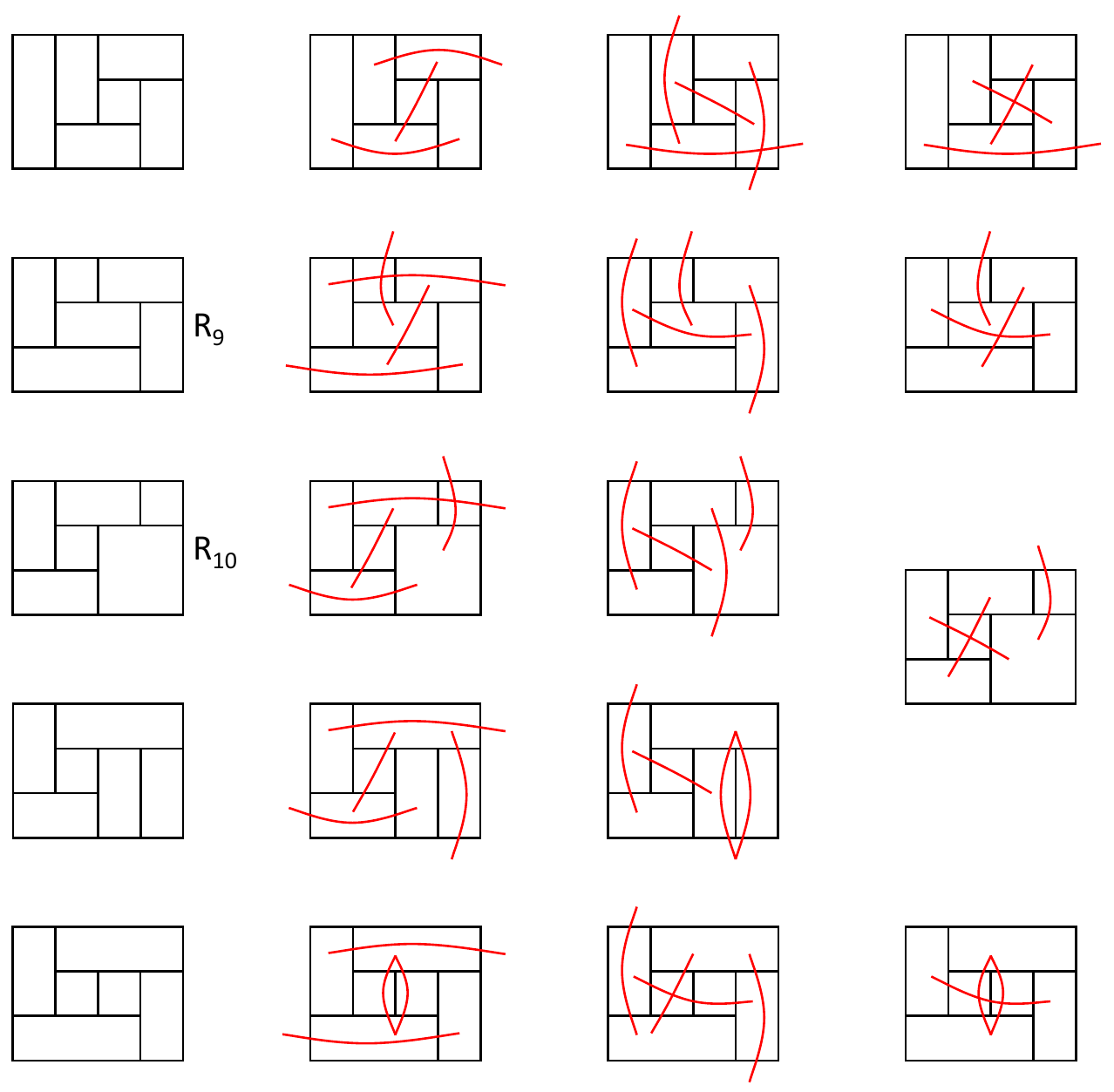}
\caption{Topologies containing the S-pattern at 6-loop, each of which has three choices of integrand numerator.
Due to their B-pattern, $R_9$ and $R_{10}$ also appear as derivative topologies of the cross family
containing the S-pattern.} \label{fig-45}
\end{center}
\end{figure}

\begin{figure}
\begin{center}
\includegraphics[width=0.86\textwidth]{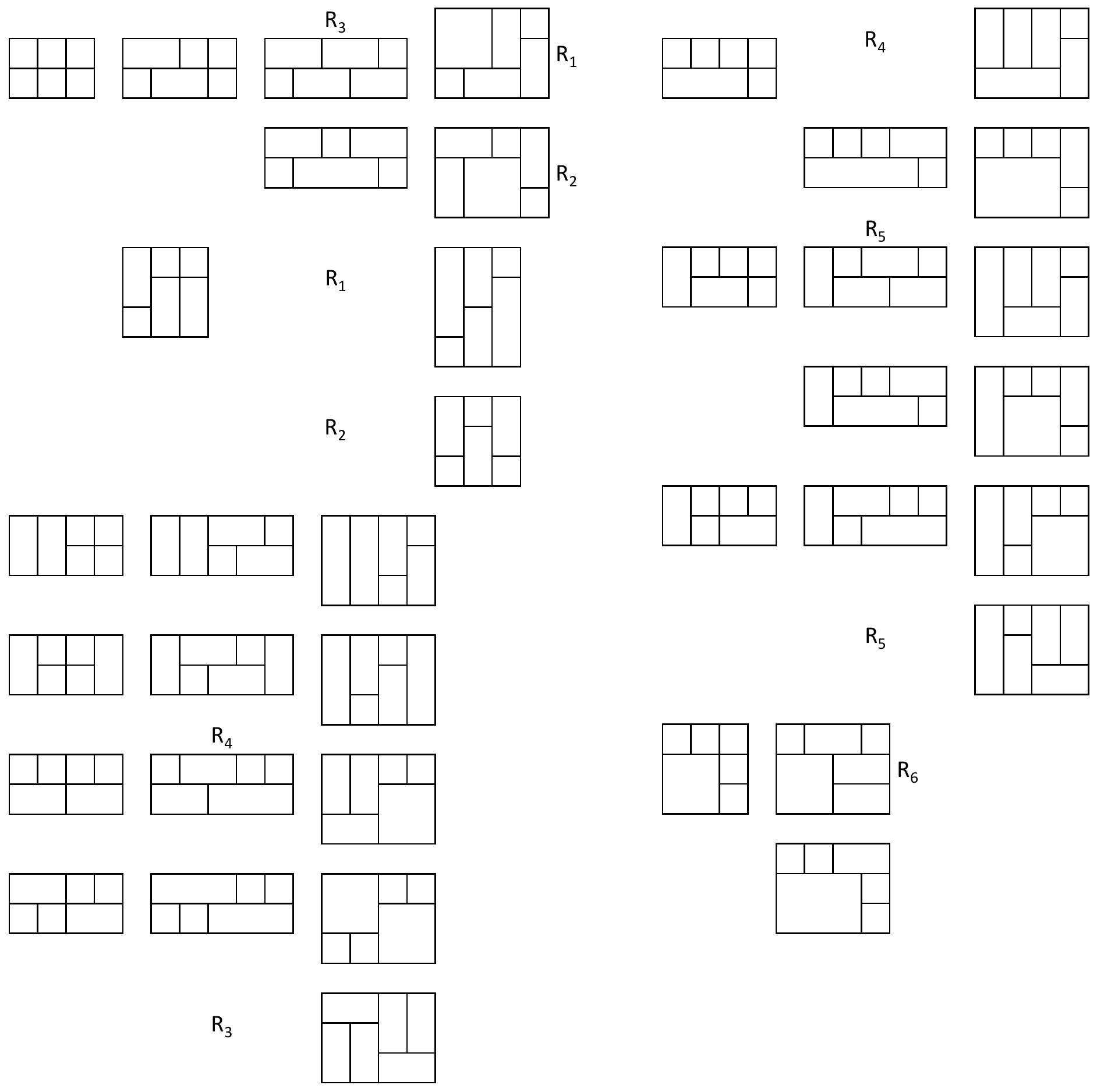}
\caption{Topology families containing the C- and B-pattern at 6-loop: part 1/2.
The 1st one in the left column is the only double-cross family.} \label{fig-46}
\end{center}
\end{figure}

\begin{figure}
\begin{center}
\includegraphics[width=0.67\textwidth]{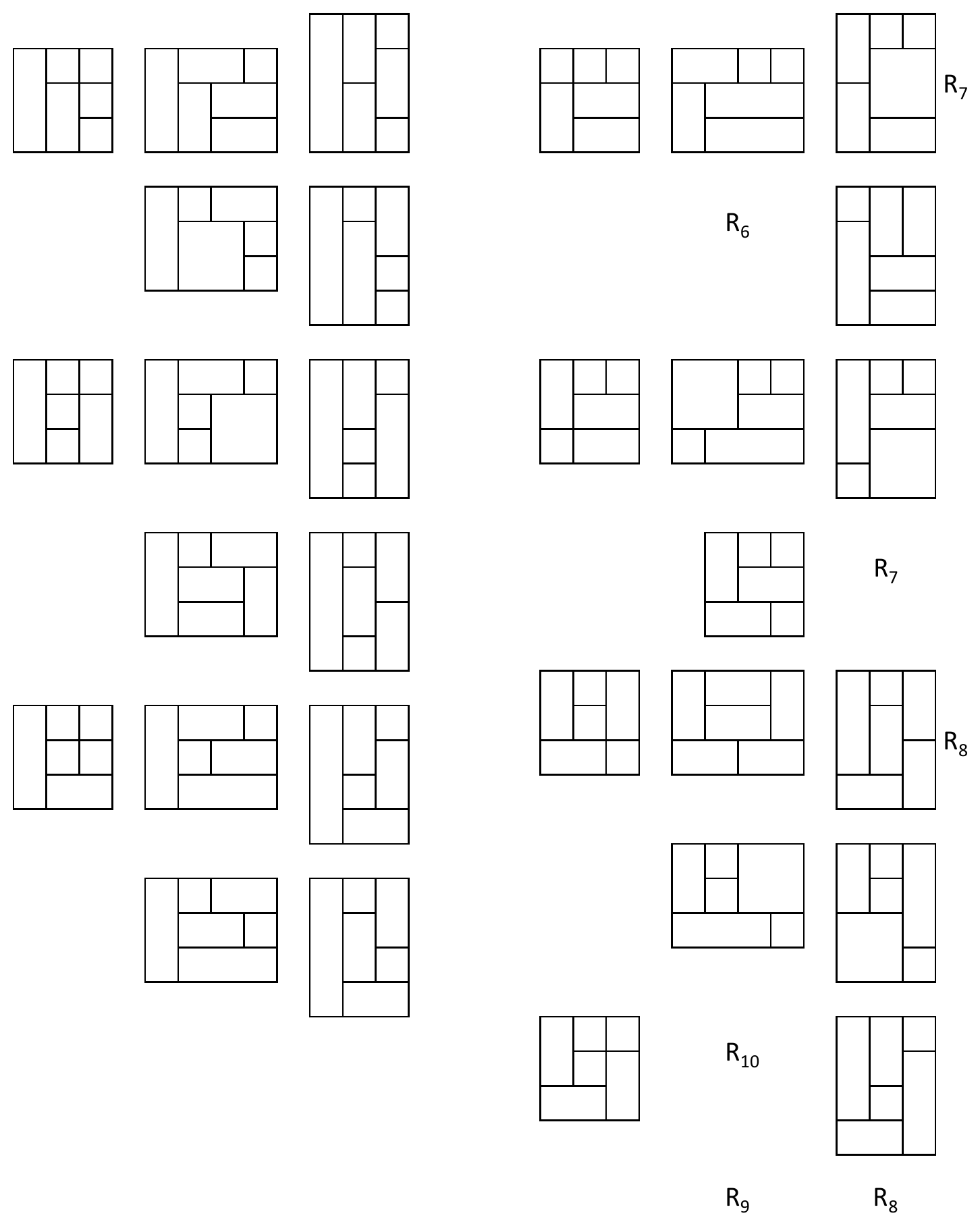}
\caption{Topology families containing the C- and B-pattern at 6-loop: part 2/2.
The last one is the only cross family containing the S-pattern.} \label{fig-47}
\end{center}
\end{figure}

Below we list all Mondrian topologies at 6-loop, with the classification specifying the C-, B- and S-pattern
as those containing none of them are classified as topologies of the L-pattern,
in figures \ref{fig-44}, \ref{fig-45}, \ref{fig-46} and \ref{fig-47}.

In figure \ref{fig-45}, each of the five topologies
containing the S-pattern has three choices of dual conformally invariant integrand numerator, which is the
straightforward generalization of \eqref{eq-10} and \eqref{eq-11} at 5-loop, as discussed in section \ref{sec3},
and don't forget the minus sign for the third choice. The third and fourth topologies share the same type of integrand
for the third choice, hence there is only one distinct diagram for both of them
(the red curves indicate numerators in terms of zone variables as usual).
In addition, the second and third topologies also contain the B-pattern, so they also appear as derivative topologies
of the only cross family that contains the S-pattern, denoted by $R_9$ and $R_{10}$ in figure \ref{fig-47}.

In figure \ref{fig-46} and \ref{fig-47}, topologies containing the C- and B-pattern are grouped
as families derived from those containing maximal numbers of crosses, where the 1st one in the left column of
figure \ref{fig-46} is the only double-cross family. In addition, topologies containing the B-pattern which appear
more than once will be denoted by $R_i$ when they first appear, while for the second time, they will be
replaced by $R_i$ only to avoid unnecessary repetition ($R$ stands for repetition).

\newpage

\end{document}